\documentclass[useAMS,usenatbib,usegraphicx]{mn2e}
\usepackage{txfonts}
\title[Insights into neutron star accretion modes]{Contrasting behaviour from
two Be/X-ray binary pulsars: insights into differing neutron star accretion
modes}
\author[L. J. Townsend et al.]{L. J. Townsend$^{1}$\thanks{E-mail:
ljt203@soton.ac.uk}, S. P. Drave$^{1}$, A. B. Hill$^{1,2}$, M. J. Coe$^{1}$, R.
H. D. Corbet$^{3}$, A. J. Bird$^{1}$\\
$^{1}$Faculty of Physical and Applied Sciences, University of Southampton,
Highfield, Southampton, SO17 1BJ, United Kingdom\\
$^{2}$W. W. Hansen Experimental Physics Laboratory, Kavli Institute for Particle
Astrophysics and Cosmology, \\ Department of Physics and SLAC National
Accelerator Laboratory, Stanford University, Stanford, CA 94305, USA\\
$^{3}$University of Maryland Baltimore County, X-ray Astrophysics Laboratory,
Mail Code 662; NASA Goddard Space Flight Center, Greenbelt, MD 20771, USA\\}

\begin{document}

\date{Accepted 2013 April 15.  Received 2013 April 4; in original form 2012 September 14.}

\pagerange{\pageref{firstpage}--\pageref{lastpage}} \pubyear{2013}

\maketitle

\label{firstpage}

\begin{abstract}

\noindent In this paper we present the identification of two periodic X-ray
signals coming from the direction of the Small Magellanic Cloud (SMC). On
detection with the \textit{Rossi X-ray Timing Explorer (RXTE)}, the 175.4\,s and
85.4\,s pulsations were considered to originate from new Be/X-ray binary (BeXRB)
pulsars with unknown locations. Using rapid follow-up \textit{INTEGRAL} and
\textit{XMM-Newton} observations, we show the first pulsar (designated SXP175)
to be coincident with a candidate high-mass X-ray binary (HMXB) in the northern
bar region of the SMC undergoing a small Type II outburst. The orbital period
(87d) and spectral class (B0-B0.5IIIe) of this system are determined and
presented here for the first time. The second pulsar is shown not to be new at
all, but is consistent with being SXP91.1 - a pulsar discovered at the very
beginning of the 13 year long \textit{RXTE} key monitoring programme of the SMC.
Whilst it is theoretically possible for accreting neutron stars to change spin
period so dramatically over such a short time, the X-ray and optical data
available for this source suggest this spin-up is continuous during long phases
of X-ray quiescence, where accretion driven spin-up of the neutron star should
be minimal.

\end{abstract}


\section{Introduction}

High-mass X-ray binaries (HMXB) consist of a massive early type main sequence or
supergiant star and a neutron star or black hole compact object. They are found
in regions of high gas and dust, such as the spiral arms of the Milky Way. As
such, they are excellent tracers of star formation and ideal objects through
which to study star forming galaxies. The main sub-groups of HMXBs are the
Be/X-ray binaries (BeXRB) and supergiant X-ray binaries (SGXB), making up a
large fraction of the known population. The newly emerging classes of supergiant
fast X-ray transients (SFXT, e.g. \citealt{sgu05}, \citealt{sid11}) and
gamma-ray binaries (e.g. \citealt{hill10}) complete the HMXB family. The
Galactic population includes all of the above classes of HMXB, as well as a
mixture of neutron star and black hole compact objects. This is markedly
different to the population in the Small Magellanic Cloud (SMC), which consists
only of BeXRBs and one SGXB, all of which have a neutron star accretor. The
number of HMXBs in the SMC is also at odds with what is known from observations
of the Galaxy. A simple mass comparison suggests there should be only one or two
such systems in the SMC, and yet there are over 60 now known. Population
synthesis models by \citet{dray06} predict the low metallicity environment in
the SMC could increase the number of HMXBs by a factor of 3. However, this is
still not sufficient to explain the SMC population on its own. A recent episode
of star formation, possibly due to an increase in the tidal force exerted on the
SMC by a close approach with the Large Magellanic Cloud (LMC; \citealt{gard96},
\citealt{diaz11}), is the current favoured scenario to explain the number of
HMXBs observed.

The most numerous of the HMXBs, the BeXRBs, are the subject of this paper. They
consist of a neutron star and an early type Be or late Oe main sequence
secondary star. The neutron star is usually in a wide, eccentric orbit allowing
for long periods of little X-ray activity. Classically, X-ray outbursts from
these systems fall into one of two types: Type I outbursts occur near periastron
passage of the neutron star where there is more accretable material, last for a
few days to a week and have luminosities around $10^{36}$\,ergs\,s$^{-1}$.
Studying these outbursts allows orbital periods to be derived through timing
analysis of their light curves. Type II outbursts are typically an order of
magnitude brighter than Type I outbursts and can last much longer. The cause of
this being an enlargement of the circumstellar material or radially driven wind
around the companion star (e.g. \citealt{neg00}), allowing accretion to occur at
any orbital phase and at a higher rate.

\begin{figure*}
\centering
 \includegraphics[width=\textwidth]{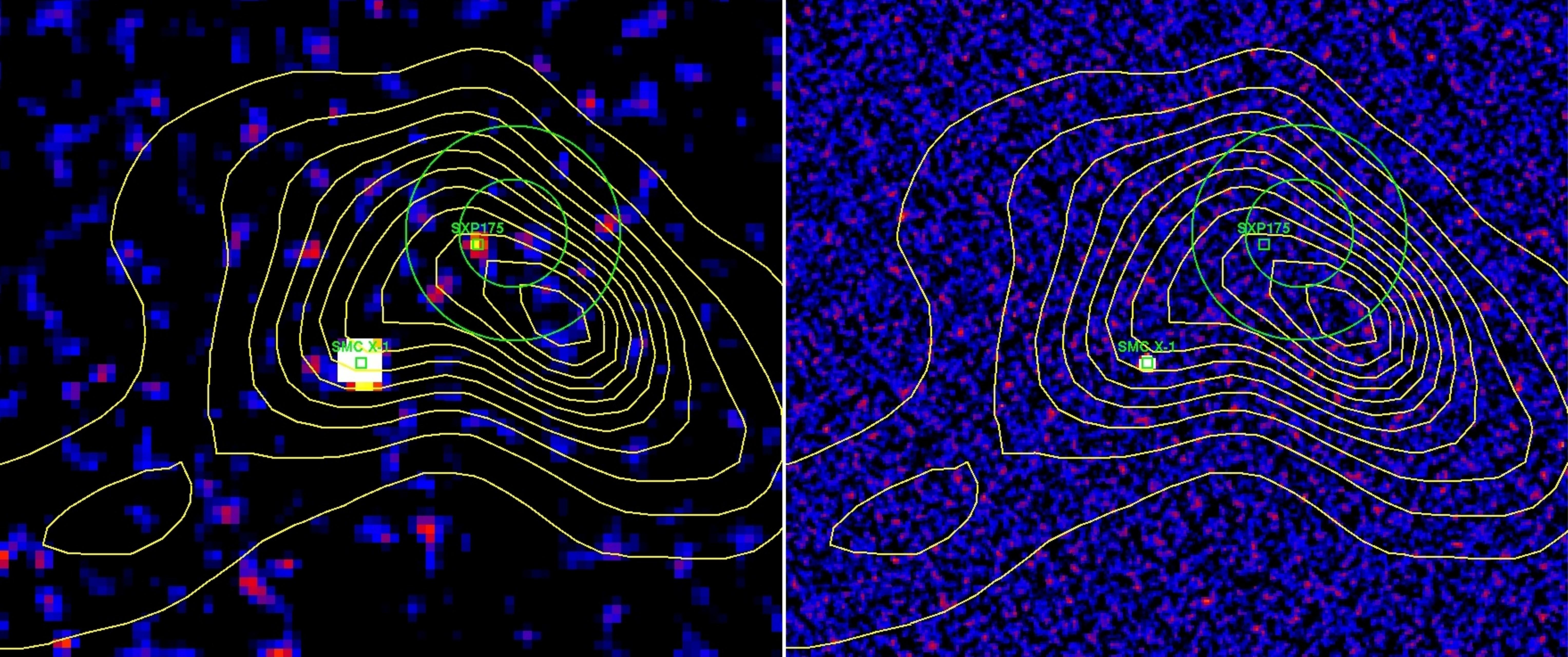}
 \caption{Left: \textit{INTEGRAL}/IBIS 15$-$60\,keV significance map of an
observation performed on 2011 April 2 (MJD 55653). A single source is detected
within the \textit{RXTE}/PCA FOV at a significance of 6.2$\sigma$. The circles
indicate the \textit{RXTE}/PCA half and zero response of the pointing used
whilst the contours are those of the HI column density of \citet{putman2003}.
Right: \textit{INTEGRAL}/JEM-X 3$-$10\,keV significance map showing a
non-detection at the location of the IBIS source.}
 \label{maps_sxp175}
\end{figure*}

The SMC provides an excellent laboratory to study both the fundamental physics
of individual BeXRB systems and the global properties of a substantial
population formed at the same time and at a well defined distance. The latter
can give an insight into the star-forming history of the galaxy and improve our
understanding of how differences in the local environment (e.g. metallicity)
affect the formation and evolution of the binary system. The former allows us to
observe the interaction between a compact object and a massive star and how the
compact object behaves under varying levels of accretion. In turn, this can
provide information about the fundamental properties of the compact object such
as the neutron star equation of state or black hole masses. In this paper, we
present X-ray and optical data of two BeXRB pulsars in the SMC during recent
X-ray outbursts. The first is a previously undiscovered system, adding to the
already vast population. We show that this system is typical of many others in
the SMC. The second system, however, shows a constant spin-up of the neutron
star during luminosity changes of several orders of magnitude. This
characteristic has not been observed before and suggests that something unusual
is happening to the accretion mode or near the emission region in this system.
The X-ray and optical data are presented in sections 2 and 3 respectively. We
discuss the data and their implications in section 4 before concluding on our
results in section 5.

\section{X-ray observations}

\subsection{SXP175 = RX\,J0101.8--7223}

The \textit{Rossi X-ray Timing Explorer (RXTE)} Proportional Counter Array (PCA)
(see \citet{jahoda06}
for instrument details and calibration model) detected $175.4\pm0.1$\,s
pulsations coming from the direction of the SMC on 2011 March 19 (MJD 55639)
during an observation as part of a key programme to monitor X-ray binary
activity in the SMC \citep{gal08}. The observation was made in the GoodXenon
event mode and the data extracted using the standard reduction commands from
HEASOFT v.6.9. The light curve was then binned at 0.01\,s before being
background subtracted and barycentre corrected. Finally, we corrected the count
rate for the number of active PCUs. The final light curve was subjected to a
Lomb-Scargle periodogram search for periodic changes in the flux. The error
associated with any period found is calculated based on the formula given in
\citet{horne86}. This period was confirmed by a second detection at
$175.1\pm0.1$\,s on 2011 March 26 (MJD 55646). It did not coincide with the spin
period of any known pulsar in the \textit{RXTE} field of view (FOV), and so was
concluded to be a new HMXB pulsar in the SMC. The entire \textit{RXTE} data
archive was subsequently searched in order to build up an activity history of
the pulsar (henceforth referred to as SXP175 under the naming convention of
\citealt{coe05}). Unfortunately, this source has shown very little X-ray
activity despite good coverage with the PCA over approximately 9 of the last 13
years. Only a handful of low significance ($<$\,2$\sigma$) detections are seen
in the long-term light curve. These detections are sporadic and no orbital
information could be drawn from the data.

An \textit{INTEGRAL} \citep{winkler03} ToO follow up observation was performed
on 2011 April 2 (MJD 55653) to localise any X-ray sources within the
\textit{RXTE} FOV. The observation consisted of 50 individual Science Windows
(ScW) summing to a total good time exposure of $\sim$89.5\,ks. Nominally, a ScW
represents an $\sim$2000\,s exposure. Both the hard IBIS/ISGRI
(\citealt{ubertini03}; \citealt{lebrun03}) and soft JEM-X \citep{lund03} X-ray
data were analysed using the \textit{INTEGRAL} Offline Science Analysis software
(OSA v9.0, \citealt{gold03}). Images were generated at ScW level in the
3$-$10\,keV and 15$-$60\,keV energy bands for JEM-X and IBIS respectively and
combined into mosaics of the total observation. These mosaics are shown in
Figure \ref{maps_sxp175} with the \textit{RXTE}/PCA pulsation detection pointing
overlaid. A single source is detected within the PCA FOV with a significance of
6.2$\sigma$ in the IBIS revolution map, with a 90\% error circle on its position
of radius 3.9 arcmin \citep{scaringi2010}. The other fluctuations visible in the
significance maps were not identified as real sources by the OSA source
detection algorithms either because they are beneath the new source detection
limit or their PSF is inconsistent with the known IBIS PSF which is well
characterised at all positions within the FOV of the instrument. The detection
limit was set at 4$\sigma$ for a source at a position coincident with a known
X-ray source and 6$\sigma$ for a potential new source that is not coincident
with a known X-ray source. As such, fluctuations that appear by eye to be of similar magnitude may be quite
different.The source was not detected in the 3$-$10\,keV energy band in the corresponding
JEM-X map.The detection permitted an \textit{XMM-Newton} ToO observation to
identify the nature of the source.

\begin{figure}
\centering
 \includegraphics[scale=0.38,angle=270]{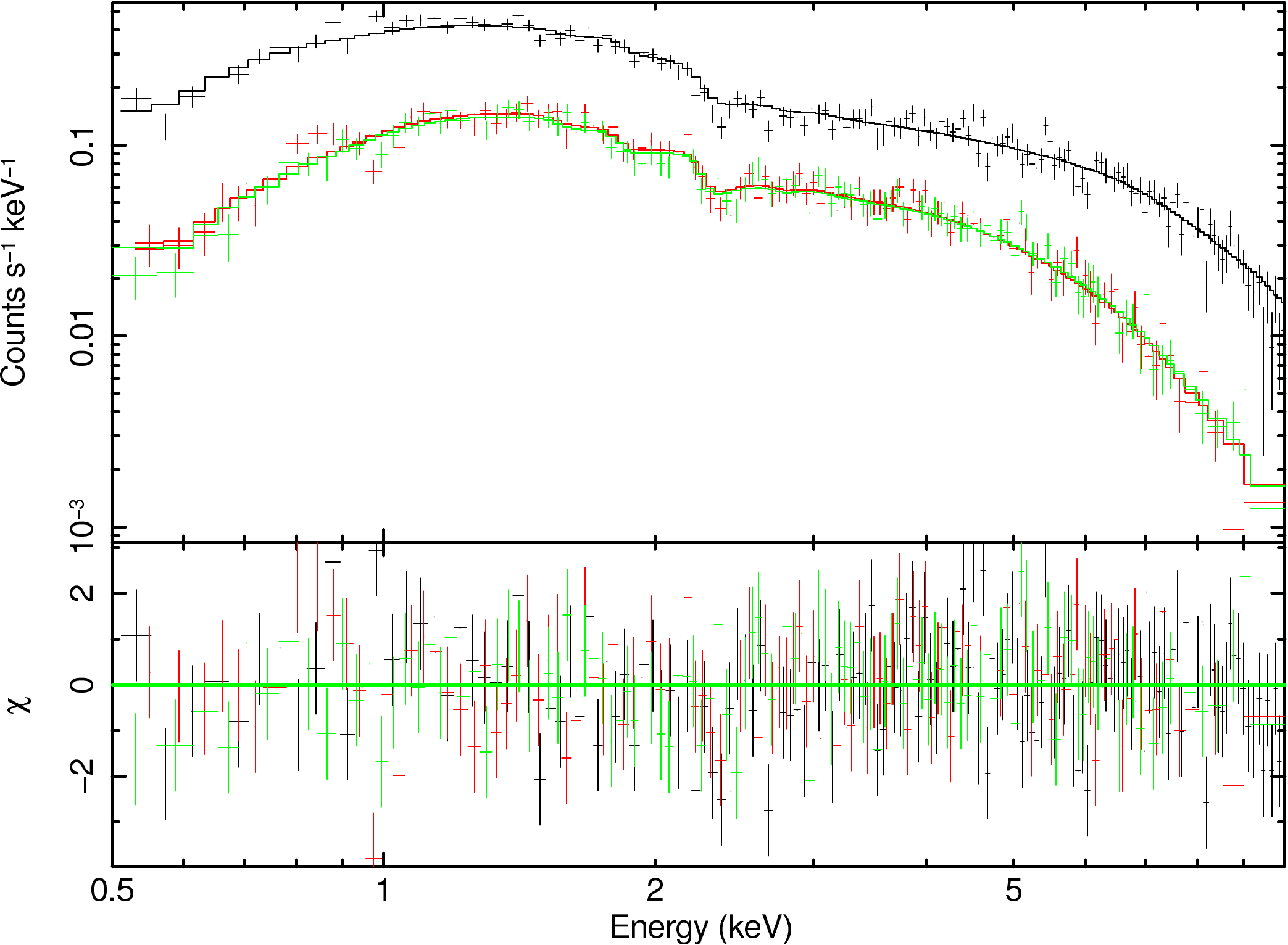}
 \caption[\textit{XMM-Newton} EPIC spectra of SXP175.]{\textit{XMM-Newton} EPIC
spectra of SXP175. EPIC-PN is shown in black and EPIC-MOS in red (MOS1) and
green (MOS2). The three spectra were fitted simultaneously with an absorbed
power law allowing only a constant normalisation factor between the three
spectra. The model fit is presented in Table \ref{tab:1}.}
 \label{fig:xmm175spec}
\end{figure}

\begin{table}
  \caption{Results of absorbed power-law fits to EPIC spectra of SXP175 and
SXP91.1.}
  \label{tab:1}
  \centering
\begin{tabular}{|l|c|c|}
\hline
  Spectral Parameter & SXP175 & SXP91.1\\
\hline
  Date & 2011/04/08 & 2009/09/27\\
  SMC \textit{N}$_{\mathrm H}$ (10$^{22}$ cm$^{-2}$) & $0.47\pm0.03$ &
$6.55\pm0.44$\\
  Photon index & $0.96\pm0.02$ & $0.95\pm0.06$\\
  $\chi^{2} _{\nu}$/degrees of freedom & 1.08/391 & 1.13/158\\
  Model flux (10$^{-12}$\,erg\,cm$^{-2}$\,s$^{-1}$)$^{(a)}$ & 8.50 (M2) & 1.92
(M2)\\
  Luminosity (10$^{36}$\,erg\,s$^{-1}$)$^{(b)}$ & 3.65 & 0.82\\
  \hline
\end{tabular}
\begin{flushleft}
$^{(a)}$\,The median 0.2--10\,keV flux from the three detectors is presented,
with the detector indicated in brackets. M2=MOS2. $^{(b)}$\,The 0.2--10\,keV
luminosity at a distance to the SMC of 60\,kpc \citep{hilditch05}. Errors
signify the 90\% confidence level.
\end{flushleft}
\end{table}

An \textit{XMM-Newton} observation was carried out on 2011 April 8 (MJD 55659)
with the EPIC cameras in full frame, imaging mode.
\textit{XMM-Newton} has three X-ray telescopes (Aschenbach et al. 2002), one
equipped with EPIC-pn (Str{\"u}der et al. 2001) and two with EPIC-MOS (Turner et
al. 2001) CCD detectors in the focal plane.
For data reduction we selected events with {\tt PATTERN$\le$12} in a circular
extraction region, placed on the source. Filtering of periods with high
background was not necessary, since soft proton flares were at a quiescent
level, yielding a net exposure time of 18.5\,ks. A background region was chosen
from a source free region of the chip in which the source was seen, such that
the RAWY positions were the same to minimise the background variation across
each chip. Source detection and light curve extraction were carried out using
standard SAS\footnote{Science Analysis Software (SAS),
http://xmm.esac.esa.int/sas/} tools. A single bright X-ray source was found
within the IBIS error circle at the position RA = 01:01:52.5, dec = -72:23:34.9
(J2000.0) with a 1$\sigma$ error circle of radius 1.1 arcsec. A Lomb-Scargle
periodogram of the 0.2--10 keV light curve revealed a period of $175.1\pm0.1$s,
confirming the \textit{RXTE}, \textit{INTEGRAL} and \textit{XMM-Newton}
detections are of the same source.

For spectral analysis we used only events with {\tt PATTERN$\le$4}. The response
matrices and ancillary files were created using the SAS tasks {\tt rmfgen} and
{\tt arfgen}. The X-ray spectrum was fit with an absorbed power law with
Galactic photoelectric absorption, N$_{\rm H}$, fixed at
$6\times10^{20}$\,cm$^{-2}$ \citep{dic90} and the SMC column density with
abundances at 0.2 for metals a free fit parameter (Figure \ref{fig:xmm175spec}).
This resulted in a photon index of $0.96\pm0.02$ and intrinsic SMC absorption of
$(4.7\pm0.3)\times10^{21}$\,cm$^{-2}$. The full list of model parameters is
presented in Table \ref{tab:1}. The 0.2--10 keV flux from the fit is
$8.5\times10^{-12}$\,ergs\,cm$^{-2}$\,s$^{-1}$, corresponding to
$3.7\times10^{36}$\,ergs\,s$^{-1}$ at the distance of the SMC (60\,kpc;
\citealt{hilditch05}). These spectral parameters are fairly standard for a BeXRB
and suggest that the non-detection with JEM-X is more likely due to a low
signal-to-noise ratio in the less sensitive of the two \textit{INTEGRAL}
instruments, rather than the hardness of the emission. The pulsar is coincident
with the HMXB candidate RX J0101.8-7223 (\citealt{hab00}, \citealt{yoko03}) and
is believed to be the same object. The counterpart is likely the emission line
star [MA93] 1288 \citep{ma93}. Optical and IR data of this star are explored in
the next section.

\begin{figure*}
 \includegraphics[width=\textwidth]{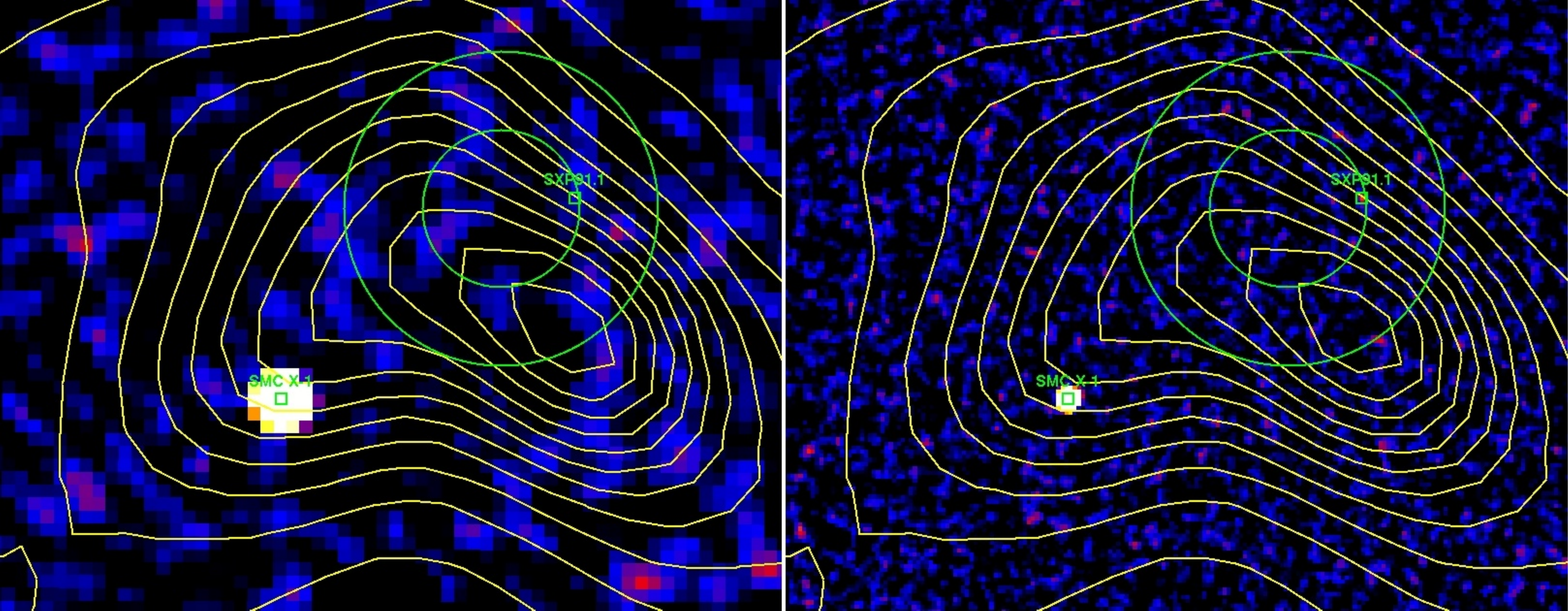}
 \caption{Left: \textit{INTEGRAL}/IBIS 15$-$60\,keV significance map of an
observation performed on 2011 May 20 (MJD 55701). No significant sources are
detected within the \textit{RXTE}/PCA FOV. As before, the circles indicate the
\textit{RXTE}/PCA half and zero response of the pointing used whilst the
contours are those of the HI column density of \citealt{putman2003}. Right:
\textit{INTEGRAL}/JEM-X 3$-$10\,keV significance map. The JEM-X source locator
tool (j\_ima\_src\_locator) identified a single source within the
\textit{RXTE}/PCA FOV detected at a significance of 5.6$\sigma$.}
 \label{maps_sxp911}
\end{figure*}

\subsection{SXP91.1 = AX\,J0051--722 = RX\,J0051.3--7216}

\begin{figure*}
\hspace*{10pt}
 \includegraphics[scale=0.7,angle=90]{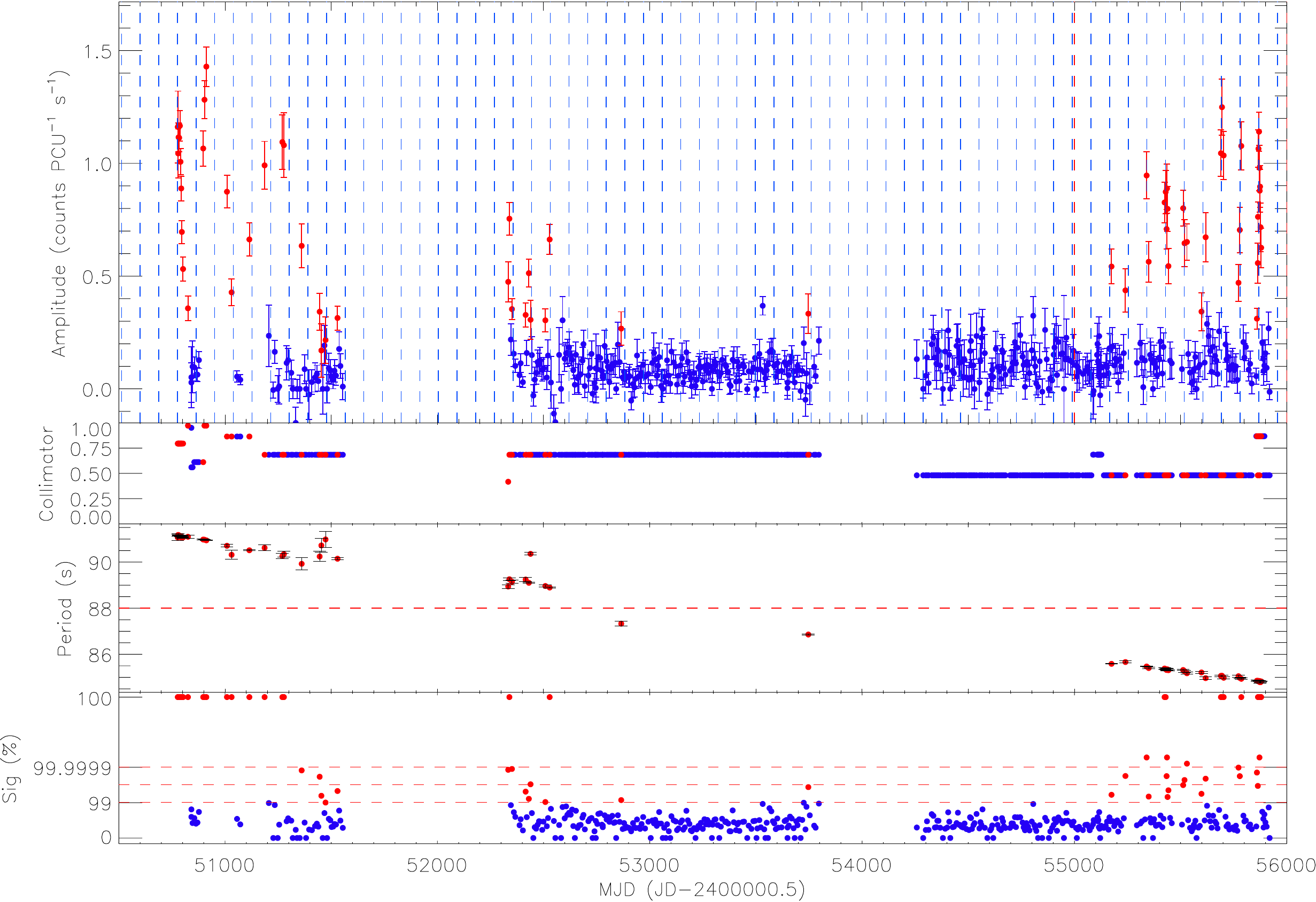}
  \caption[Long-term \textit{RXTE}/PCA light curve of SXP91.1 in the 3--10\,keV
band.]{Long-term \textit{RXTE}/PCA light curve of SXP91.1 in the 3--10\,keV
band. The top panel shows the amplitude of the pulsed emission, where each point
is a single observation. The vertical blue dashed lines show the most likely
X-ray ephemeris based on an orbital period search of the light curve. The second
panel shows the source position within the \textit{RXTE} FOV. A source at the
very centre has a collimator response of 1 and a source at the very edge 0. The
collimator response is approximately linear with distance from the centre. The
third panel shows the pulse period measurements in seconds. The final panel
shows the significance of the detection. The calculation of the significance is
described in the text. Detections above a significance level of 99\% are plotted
in red, with anything below this level plotted in blue. The source had good
coverage with \textit{RXTE} between MJD 52300--55000, though very few X-ray
outbursts were detected. A constant spin-up can be fit to the entire spin period
history, despite this apparent lack of accretion. This is discussed in greater
detail in the text.}
 \label{fig:sxp91.1hist}
\end{figure*}

SXP91.1 = AX\,J0051--722 = RX\,J0051.3--7216 was the first pulsar to be
discovered in the \textit{RXTE} monitoring programme at a period of
$92\pm1.5$\,s \citep{marshall97}. Further analysis refined this measurement to
$91.12\pm0.05$\,s \citep{cor98}. The original positional uncertainty from the
\textit{ASCA} observation was 1.2 arcmin, however, this was later improved on by
a \textit{ROSAT} observation that yielded a position of RA = 00:50:59.3, dec =
-72:13:26 (J2000.0) with a positional uncertainty of radius 2.6 arcsec
\citep{sasaki00}. The source was regularly seen during the first 2 years of the
programme (1998-9) and again briefly during the fifth year of the programme
(2002). After that it was not detected again with any certainty for over 7 years
despite good coverage with \textit{RXTE} for most of this period.

\textit{RXTE} observations performed on 2010 August 16 and 2010 August 21 showed
the presence of pulsations with a period of $85.4\pm0.1$\,s in their power
spectra \citep{cor10}. On seeing this pulsation, it was believed not to be
consistent with any known pulsar in the SMC. The period is close to the second
harmonic of the known pulsars SXP169 and SXP172, but neither appeared likely to
be the source of the 85.4\,s pulsations. Recent measurements of SXP169 have
shown a spin-up trend that has produced a period significantly different from
twice 85.4\,s (e.g. \citealt{gal08}) and the detection times are not consistent
with the known orbital period or ephemeris of SXP169. The collimator response to
SXP172 in the pointed observation is of the order 5\%, making this source
unlikely to be the origin of the pulsations unless it was accreting close to the
Eddington limit for a 1.4\,M$_{\odot}$ neutron star. In addition, we see no
evidence for modulation near twice 85.4\,s in the power spectra of the light
curves, but the second harmonic of this period is present at 42.7\,s. This
pulsation was thus concluded to be from a new HMXB which was designated SXP85.4
\citep{cor10}.

A Lomb-Scargle periodogram of the X-ray light curve allowed an epoch of maximum
pulsed flux to be calculated. An orbital period of $88.42\pm0.14$\,d
(90\% confidence) is the best estimate derived from these X-ray data.
From this ephemeris, the time of the next outburst was predicted to allow
scheduling of an \textit{INTEGRAL} follow-up observation to try and localise the
X-ray pulses. An \textit{INTEGRAL} ToO observation was carried out on 2011 May
20 (MJD 55701). The data were processed using the same methods and energy bands
as the SXP175 observations. No sources were detected within the
\textit{RXTE}/PCA FOV by IBIS (Figure \ref{maps_sxp911}, left). However the
JEM-X source locator tool (j\_ima\_src\_locator) identified a clear source in
the JEM-X combined mosaic at RA = 00:51:00, dec = -72:14:20 (2000) with a
significance of 5.6$\sigma$ and an uncertainty of 3 arcmin (Figure
\ref{maps_sxp911}, right). Along with SMC X$-$1, this was the only X-ray source
identified by the source locator tool within the image. The position calculated
from the JEM-X mosaic is completely consistent with that of SXP91.1. From this
realisation, we immediately combined the outburst histories of SXP85.4 and
SXP91.1. The result is plotted in Figure \ref{fig:sxp91.1hist}. A description of
each panel is given in the figure caption. The errors on the spin period
measurements are calculated as stated earlier and only detections of periodic
modulation above a significance of 99\% are plotted for clarity reasons. The
significance of each detection of a given frequency is related to its
Lomb-Scargle power, P, by the following formula:

\begin{eqnarray}
 \mathrm{significance} = 100 \times \left(1-e^{-P}\right)^{M}
  \label{equ:sig}
\end{eqnarray}

\bigskip
\noindent where M is the number of independent frequencies and is typically
$2\times10^{5}$ in our analysis pipeline. As can be seen, the spin period
evolution of the recent episode of Type I outbursts is consistent with the
evolution of the period between MJD 50800--51600. Along with the position
determined by \textit{INTEGRAL}, this is strong evidence that SXP85.4 is
actually SXP91.1. Thus, this source will herein be discussed as SXP91.1. An even
more intriguing observation is that SXP91.1 appears to be spinning up at a
constant rate, through long periods of X-ray emission and through long periods
of X-ray quiescence. This unexpected result could have significant implications
for the accretion mode or the environment around the emission region in this
system and is discussed in detail later in the paper. By analysing the data
between MJD 55000--56000 we determine the orbital period and ephemeris of
maximum amplitude of this system to be 87.80$\pm$0.11\,d and MJD 55429.0$\pm$2.6
respectively (90\% confidence). The reason for the small discrepancy
between this value and that stated earlier ($88.42\pm0.14$\,d) is that the data
were re-analysed to better remove the effect of a strong outburst from another
pulsar in the power spectrum. This resulted in a spin period measurement being
excluded from the more recent orbital period analysis as it was the product of a
noisy power spectrum. A Lomb-Scargle analysis of the entire light curve
reproduces these results at a lower level of significance. This is because the
apparent Type II outburst around MJD 50800 and the sparseness of the
observations during the early years of the programme detected fewer periodic
outbursts than the most recent years of data. These results will be compared to
those found from the analysis of SXP91.1s optical light curve and discussed
further in the coming sections.

 A search through the \textit{XMM-Newton} archive showed two observations were
previously made containing the position of SXP91.1 (observations 0301170201 and
0601210701), taken on MJD 53817.3 and 55101.7 respectively. A similar search of
the \textit{Chandra}, \textit{ROSAT} and \textit{Swift} archives did not return
any useful observations. The latter of the two \textit{XMM-Newton} observations
was taken just before \textit{RXTE} detected the 85.4\,s pulsations, though no
periodicities were found in the data. A spectrum was extracted for the PN and
MOS2 detectors as described for SXP175 (the source was not inside the MOS1
detector FOV). These spectra were fit simultaneously with parameters fixed as
described earlier. This resulted in the list of model parameters presented in
Table \ref{tab:1}. The 0.2--10 keV flux from the fit is
$1.9\times10^{-12}$\,ergs\,cm$^{-2}$\,s$^{-1}$, corresponding to
$8.2\times10^{35}$\,ergs\,s$^{-1}$ at the distance of the SMC. The source was
not observed to be active in the other \textit{XMM-Newton} observation. An upper
limit for the flux was derived, corresponding to a limiting luminosity of
$\sim$\,$4\times10^{33}$\,ergs\,s$^{-1}$ (Haberl 2011, priv. comm.). This upper
limit confirms that the source was indeed in deep quiescence during the period
of non-detections with \textit{RXTE}. These observations raise two important
questions about the mode of accretion in this system: How has the neutron star
spun up during such long periods of quiescence and why are pulsations not
detected at a luminosity of $8.2\times10^{35}$\,ergs\,s$^{-1}$, typical of the
Type I outburst luminosity of almost every other BeXRB system?

\begin{figure}
\centering
 \includegraphics[scale=0.38,angle=270]{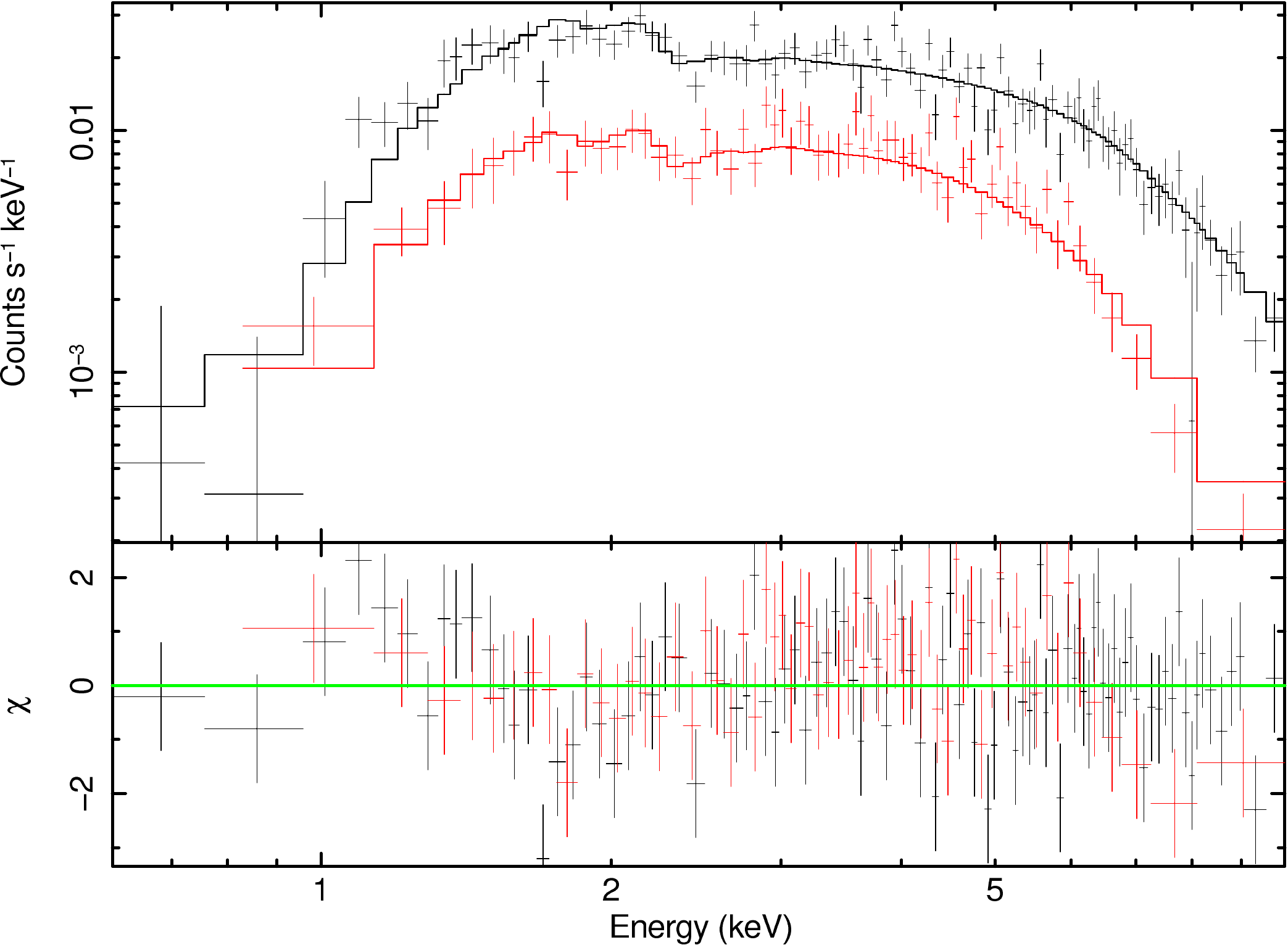}
 \caption{\textit{XMM-Newton} EPIC spectra of SXP91.1. EPIC-PN is shown in black
and EPIC-MOS2 in red. The two spectra were fitted simultaneously with
an absorbed power law allowing only a constant normalisation factor between the
two spectra. The model fit is presented in Table \ref{tab:1}.}
 \label{fig:xmm91.1spec}
\end{figure}

\section{Optical and Infrared observations}

In this section we present an optical light curve, near-IR fluxes and optical
spectra of the optical counterpart to SXP175 with the aim of understanding more
fully the circumstellar disc and its role in the X-ray outburst and to
spectrally classify the optical counterpart. We also present an optical light
curve and spectra of the counterpart to SXP91.1 to try and understand what role
the stellar environment has played in producing the large variations in X-ray
flux seen from this system during the last 14 years.

\subsection{SXP175}

\begin{figure}
 \includegraphics[scale=0.34,angle=90]{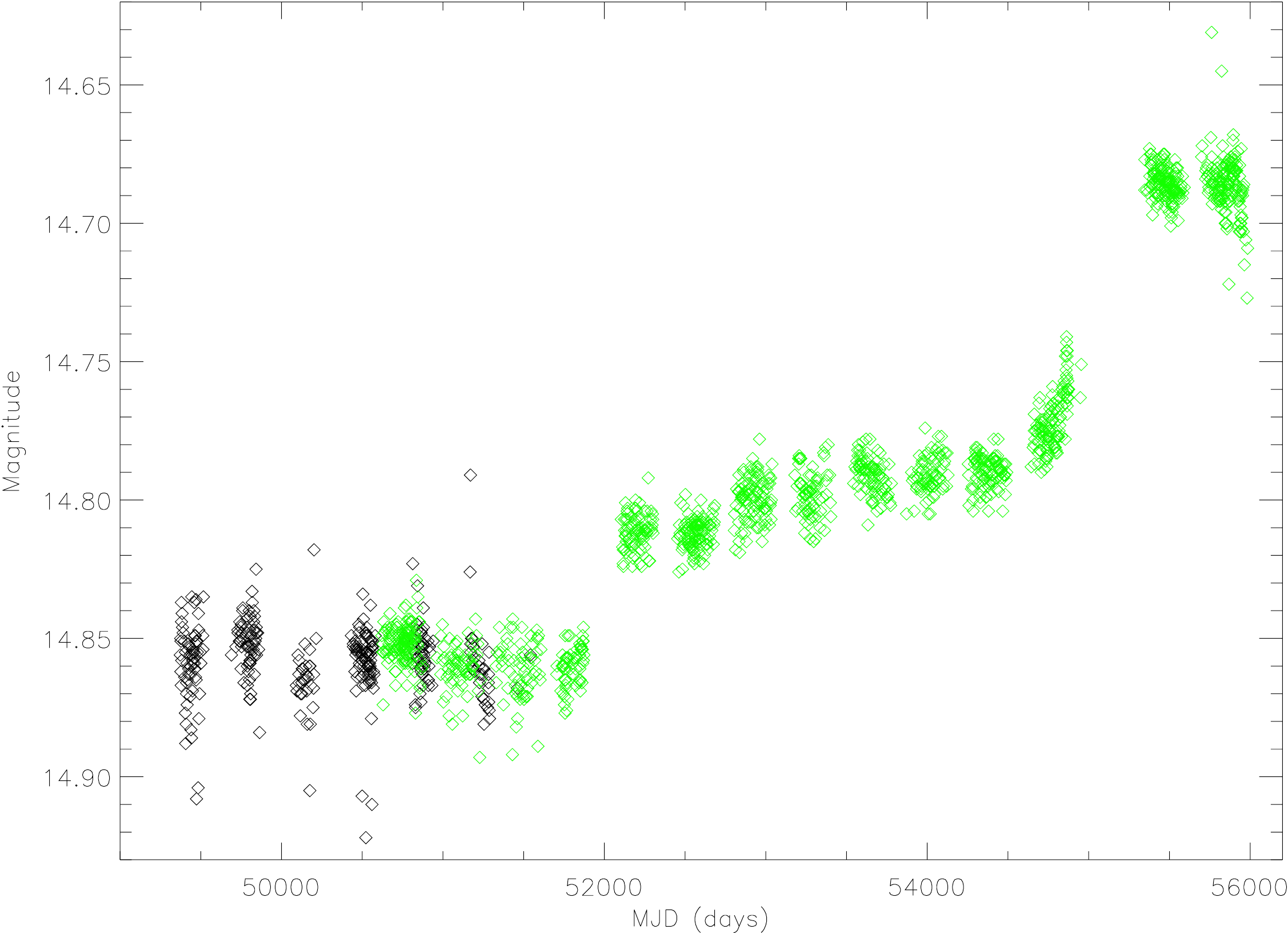}
 \caption{MACHO R-band (black points) and OGLE I-band (green points) light curve
of the optical counterpart to SXP175. The MACHO data have been arbitrarily
shifted to align roughly with the OGLE magnitude measurements at that time. The
jumps at around MJD 52000 and MJD 55000 are the times at which the OGLE project
moved from its second to third and third to fourth phases respectively (OGLE II,
III \& IV). As such, we cannot be sure whether these increases in flux are real
or whether they are a calibration effect.}
 \label{fig:175optlc}
\end{figure}

Figure \ref{fig:175optlc} shows the combined MACHO R-band and OGLE (Optical
Gravitational Lensing Experiment) I-band light curve of the optical counterpart
to SXP175. The star is \#1288 in the catalogue of \citet{ma93} and has the MACHO
I.D. 207.16716.3 and OGLE II I.D. SMC-SC9 78833\footnote{Light curves of objects
associated with variable X-ray sources can be found in the OGLE real time
monitoring of X-ray variables (XROM) programme. See
http://ogle.astrouw.edu.pl/ogle3/xrom/xrom.html and \citet{uda08} for details.}.
It shows very little periodic or long-term variability, besides a small increase
in brightness around MJD 54800 by $\sim$\,0.05 magnitudes. At this time, there
was good coverage of the source position with \textit{RXTE}, though no X-ray
activity was seen. This suggests the circumstellar disc had grown in the time
between the end of the OGLE III light curve and the X-ray outburst. On three
occasions near-infrared (NIR) data of the counterpart were obtained using the
1.4\,m Infrared Survey Facility (IRSF) situated at the South African
Astronomical Observatory (SAAO) to monitor the disc activity. The IRSF is a
Japanese built telescope designed specifically to take simultaneous photometric
data in the J, H \& $K_{s}$ bands with the SIRIUS (Simultaneous three-colour
InfraRed Imager for Unbiased Survey) camera \citep{nag99}. Data reduction was
performed using the dedicated SIRIUS
pipeline\footnote{
http://www.z.phys.nagoya-u.ac.jp/$\sim$nakajima/sirius/software/software.html}.
This performs the necessary dark subtraction, flat fielding, sky subtraction and
recombines the dithered images, as well as producing a photometric catalogue of
point sources in the reduced image. These data are presented in Table
\ref{ta:irsf} and confirm the very stable nature of the disc over a long period
of time. The only optical or NIR data taken closer to the X-ray outburst are the
two years of OGLE IV data on the right of Figure \ref{fig:175optlc}. As
mentioned, the absolute calibration between the phases of the OGLE programme is
uncertain and could vary by up to 0.2 magnitudes. As such, we cannot put any
trust in the leap of around 0.1 magnitudes between phases III and IV. In fact,
the brightening evident at the end of phase III is not obvious in phase IV,
which is very flat and shows little sign of disc growth even though these data
coincide with the X-ray outburst.

\begin{table}
\begin{center}
  \caption{IRSF IR photometry of the counterpart to SXP175. The first
measurement comes from Kato et al (2007) using the same telescope and camera as
the rest of the data.\label{ta:irsf}}
  \begin{tabular}{@{}cccc@{}}
  \hline
   Date (MJD) & J & H & K$_{s}$\\[1ex]
  \hline
   17-09-2002 (52534) & $14.82\pm0.01$ & $14.73\pm0.02$ & $14.53\pm0.02$\\[1ex]
   18-12-2007 (54452) & $14.83\pm0.01$ & $14.67\pm0.02$ & $14.50\pm0.03$\\[1ex]
   08-12-2008 (54808) & $14.82\pm0.01$ & $14.65\pm0.01$ & $14.47\pm0.02$\\[1ex]
   14-12-2009 (55179) & $14.83\pm0.02$ & $14.65\pm0.02$ & $14.45\pm0.05$\\[1ex]
  \hline
  \end{tabular}\\
\end{center}
\end{table}

\begin{figure}
 \includegraphics[scale=0.37,angle=90]{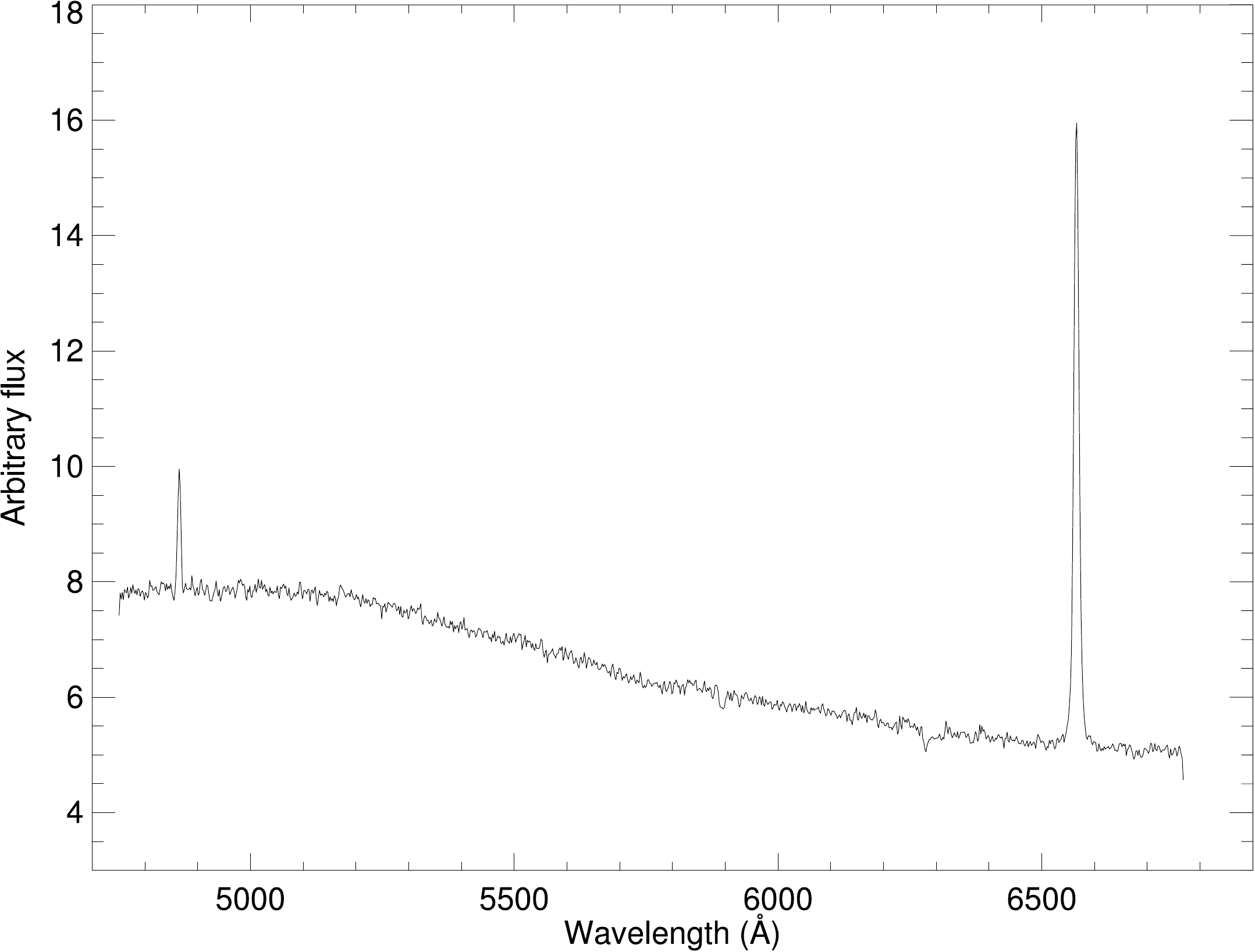}
 \caption[Optical spectrum of SXP175 taken with the 3.6\,m telescope at La
Silla.]{Optical spectrum of SXP175 taken with the ESO Faint Object Spectrograph
and Camera (EFOSC) on the 3.6\,m telescope at La Silla, Chile. H$\alpha$ and
H$\beta$ are observed to be in emission.}
 \label{fig:175spec}
\end{figure}

\begin{figure}
\hspace{10pt}
 \includegraphics[scale=0.33,angle=90]{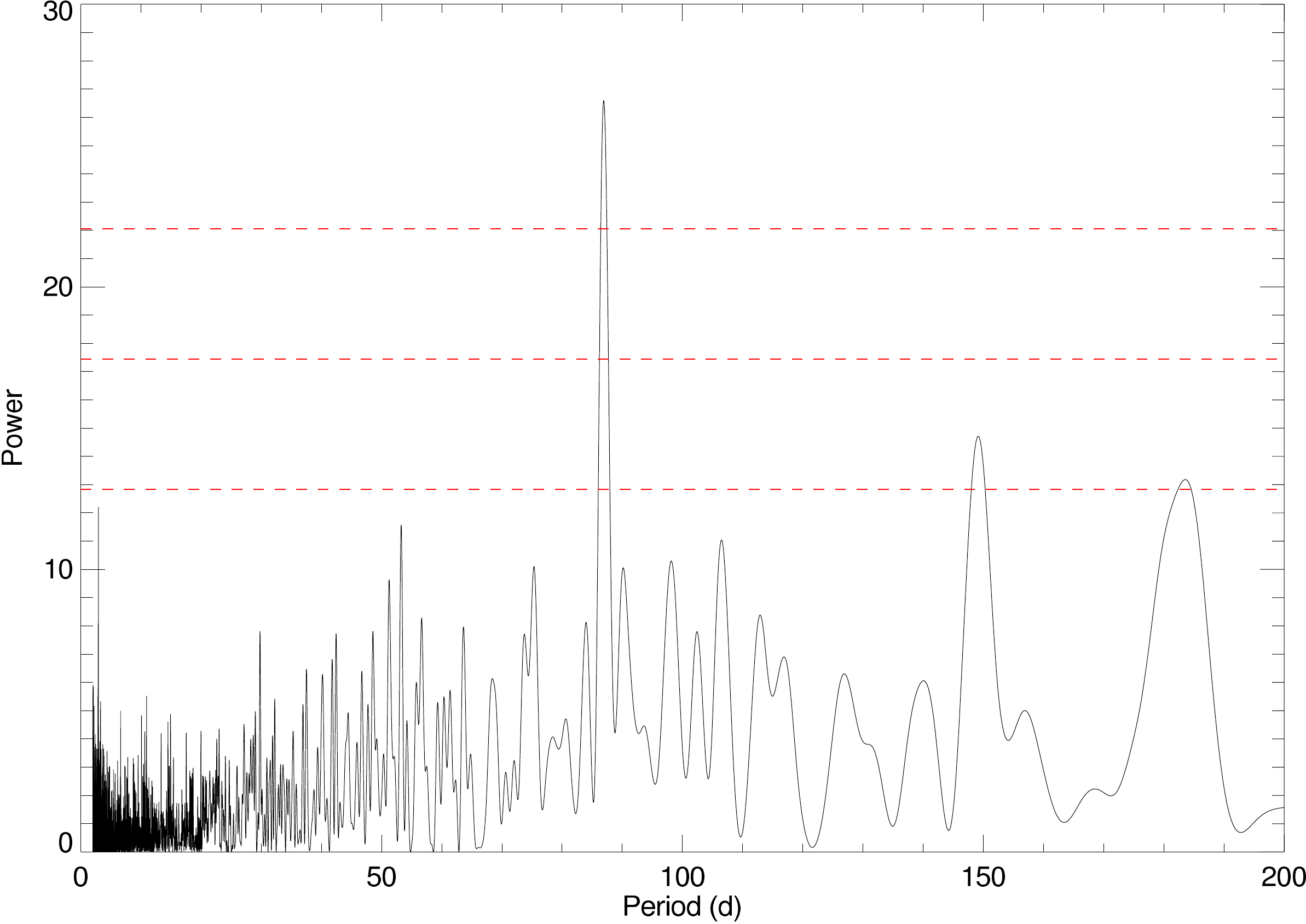}
 \caption{Power spectrum of the optical light curve of the counterpart to
SXP175. Horizontal lines are 99\%, 99.99\% and 99.9999\% confidence levels. The
peak period is $86.9\pm0.1$\,d.}
 \label{fig:175optls}
\end{figure}

\begin{figure*}
\hspace*{15pt}
 \includegraphics[scale=0.72,angle=90]{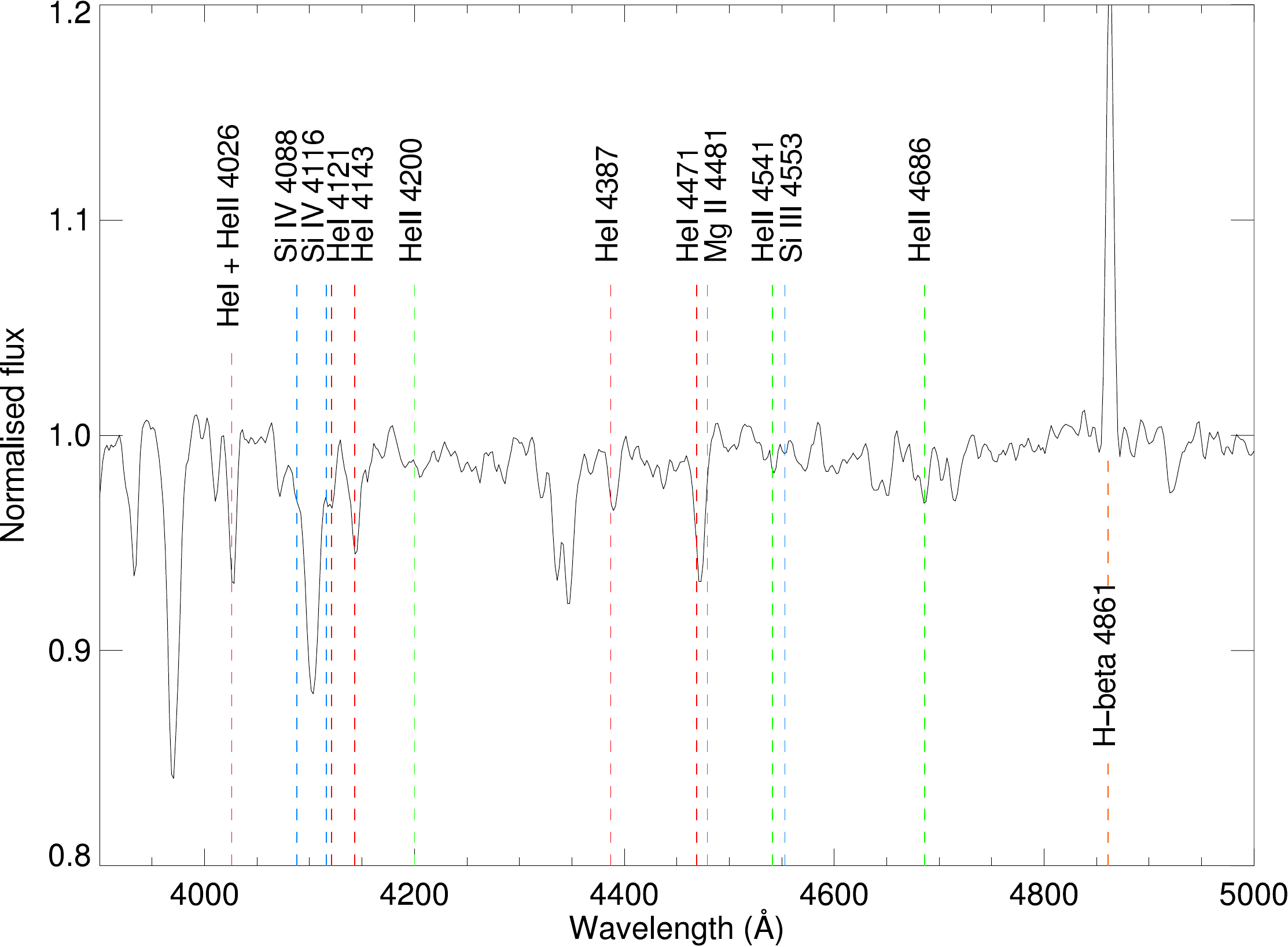}
 \caption[Spectrum of SXP175 taken with the 3.6\,m telescope at La Silla,
Chile.]{Spectrum of SXP175 taken with the EFOSC spectrograph on the 3.6\,m
telescope at La Silla, Chile. The spectrum has been normalised to remove the
continuum and the data have been redshift corrected by -150 km s$^{-1}$ to
account for the recession of the SMC. Overplotted are various atomic transitions
that are significant in the spectral classification of an early type star at
their rest wavelengths; He II, He I and metal transitions are in green, red and
blue respectively.}
 \label{fig:175class}
\end{figure*}

A red and blue spectrum of this source were taken from the ESO 3.6\,m telescope
at La Silla, Chile on 2007 September 18 and 2007 September 19.
The data were obtained with the ESO Faint Object Spectrograph (EFOSC2) mounted
at the Nasmyth B focus of the 3.6m New Technology Telescope (NTT), La Silla,
Chile. The EFOSC2 detector (CCD\#40) is a  Loral/Lesser, Thinned, AR coated, UV
flooded, MPP chip with 2048$\times$2048 pixels corresponding to
4.1\arcmin$\times$4.1\arcmin on the sky. The instrument was in longslit mode
with a slit width of 1.5\arcsec. Grisms 14 and 20 were used for blue and red end
spectroscopy respectively. Grism 14 has a a grating of 600~lines~mm$^{-1}$ and a
wavelength range of $\lambda\lambda3095$--$5085$~\AA{} ~producing a dispersion
of 1~\AA{}~pixel$^{-1}$. The resulting spectra have a spectral resolution of
$\sim12$~\AA{}. Grism 20 is one of the two new Volume-Phase Holographic grisms
recently added to EFOSC2. It has 1070 lines~pixel$^{-1}$ but a smaller
wavelength range, from 6047--7147~\AA{}. ~This results in a superior dispersion
of 0.55 \AA{}~pixel$^{-1}$ and produces a spectral resolution for our red end
spectra of $\sim6$~\AA{}.~Filter OG530 was used to block second order effects.
The data were reduced using the standard packages available in the Image
Reduction and Analysis Facility (\textsf{IRAF}). Wavelength calibration was
implemented using comparison spectra of Helium and Argon lamps taken through out
the observing run with the same instrument configuration.

In the following section we will used the blue spectrum to classify the star
based in its absorption lines and their ratios. The red end spectrum is
presented in Figure \ref{fig:175spec}, from which some important information can
be gained. Firstly, the presence of H$\alpha$ and H$\beta$ lines in emission
confirm the identification of this system as an emission line XRB. The shape of
both lines is very narrow and single peaked, suggesting the disc has a low
inclination to our line-of-sight. Secondly, the equivalent widths of the lines
can be measured. We estimate the equivalent width of the H$\alpha$ and H$\beta$
lines to be -25.7\.{A} and -1.9\.{A} respectively. This value places the system
nicely amongst other BeXRBs in the H$\alpha$ equivalent width against orbital
period distribution (\citealt{reig97}, \citealt{ant09}), and suggests the system
should have an orbital period of approximately 90--100\,d. This number is also
implied from the Corbet diagram.

To this end, the light curve in Figure \ref{fig:175optlc} was detrended and
searched for periodicities. The power spectrum is shown in Figure
\ref{fig:175optls}. A peak at $86.9\pm0.1$\,d is found at greater than the
99.9999\% confidence level. This seems likely to be the orbital period of the
system, though until there is evidence of this period in the X-ray light curve,
it cannot be certain that it is orbital in nature. Detrending was done with a
simple third order polynomial. The peak found may move around slightly with a
different detrending method, though it is safe to conclude the presence of a
significant peak in the power spectrum at $\sim$\,90\,d that is likely the
orbital period of the binary system.

\subsubsection{Spectral classification}

A star can be classified as emission line if H$\alpha$ (or any other hydrogen
line usually seen in absorption in the hot, ionised photosphere) is filled in or
is in emission, and so spectral coverage of the red part of the spectrum is
vital. However, the spectral classification of emission line stars suffers many
difficulties due to their very nature; the Balmer lines in particular will be
rotationally broadened due to the high rotational velocities of Be stars and
hence may obscure any comparisons to closely neighbouring lines, whilst the
filling-in of lines caused by the disc emission also makes classification
harder. Thus, it is beneficial to perform classification observations when the
disc emission is minimal. Further difficulties arise when observing in the
Magellanic Clouds. Classification of Galactic Be stars relies on using the ratio
of many metal-helium lines \citep{wal90}. However, this type of classification,
based on the Morgan-Keenan (MK; \citealt{mkk43}) system, is particularly
difficult in the low metallicity environment of the Magellanic Clouds because
these metal lines are either very weak or not present at all. Using high
signal-to-noise ratio spectra of SMC supergiants, \citet{len97} devised a system
for the classification of stars in the SMC that overcomes the problems with low
metallicity environments. This system is ‘normalised’ to the MK system such that
stars in both systems exhibit the same trends in their line strengths. Thus, we
have used the classification method as laid out in \citet{len97} and utilised
further in \citet{evans04}. For the luminosity classification we adopted the
classification method set out in \citet{wal90}.

The normalised and redshift corrected spectrum used to spectrally classify
SXP175 is presented in Figure \ref{fig:175class}. The spectrum shows evidence of
He II $\lambda$4686 absorption, albeit fairly weak, meaning the star must be of
type earlier the B\,1 (\citealt{len97}; \citealt{evans04}). Other ionised helium
lines (He II $\lambda\lambda$4200, 4541) may be present, but at a very weak
level. This rules out a spectral type earlier than O\,9.5. Given all of the He
II lines are weak, we would suggest a B\,0.5 class. However, if He II
$\lambda\lambda$4200, 4541 are present, the spectral type is more likely B\,0.
As it is difficult to confirm this above the noise in the continuum, we suggest
a B\,0--B\,0.5 classification. As stated previously, the lack of strong metal
lines makes the luminosity classification difficult. The He I $\lambda$4121/He I
$\lambda$4143 ratio strengthens towards more luminous stars and is the only
ratio that can be used here. The ratio in this spectrum is reasonably high,
suggesting the luminosity class may be closer to III than V. A V-band magnitude
of 14.6 \citep{hep08} allows a check of this classification by comparing the
absolute magnitude of the source with a distance modulus for the SMC of 18.9
\citep{hhh03}. Thus, an absolute magnitude of -4.3 suggests (using absolute
magnitudes for OeBe stars from \citealt{weg06}) a luminosity class of III, as a
V--IV classification would suggest a spectral type as early as O\,9. We suggest
here a classification of B\,0--B\,0.5 IIIe is the most likely for this system
given the information available. This is consistent with that of other BeXRBs in
the SMC and in the Galaxy (for example \citealt{mcbride08}; \citealt{ant09}).

\subsection{SXP91.1}

\begin{figure}
 \includegraphics[scale=0.34,angle=90]{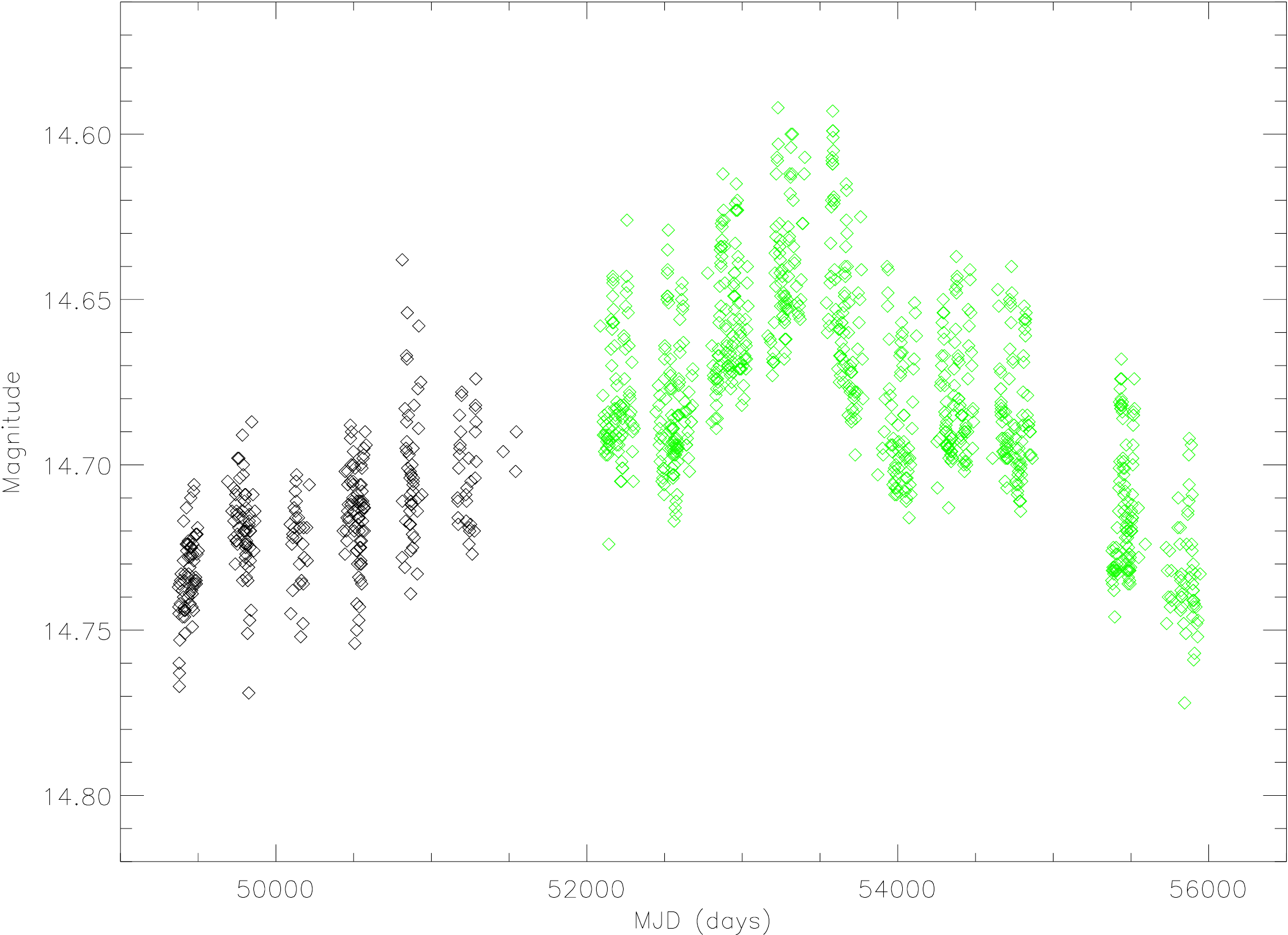}
 \caption{MACHO R-band (black points) and OGLE I-band (green points) light curve
of the optical counterpart to SXP91.1. The MACHO data have been arbitrarily
shifted to align roughly with the OGLE magnitude measurements at that time.}
 \label{fig:91.1optlc}
\end{figure}

\begin{figure}
\hspace{10pt}
 \includegraphics[scale=0.33,angle=90]{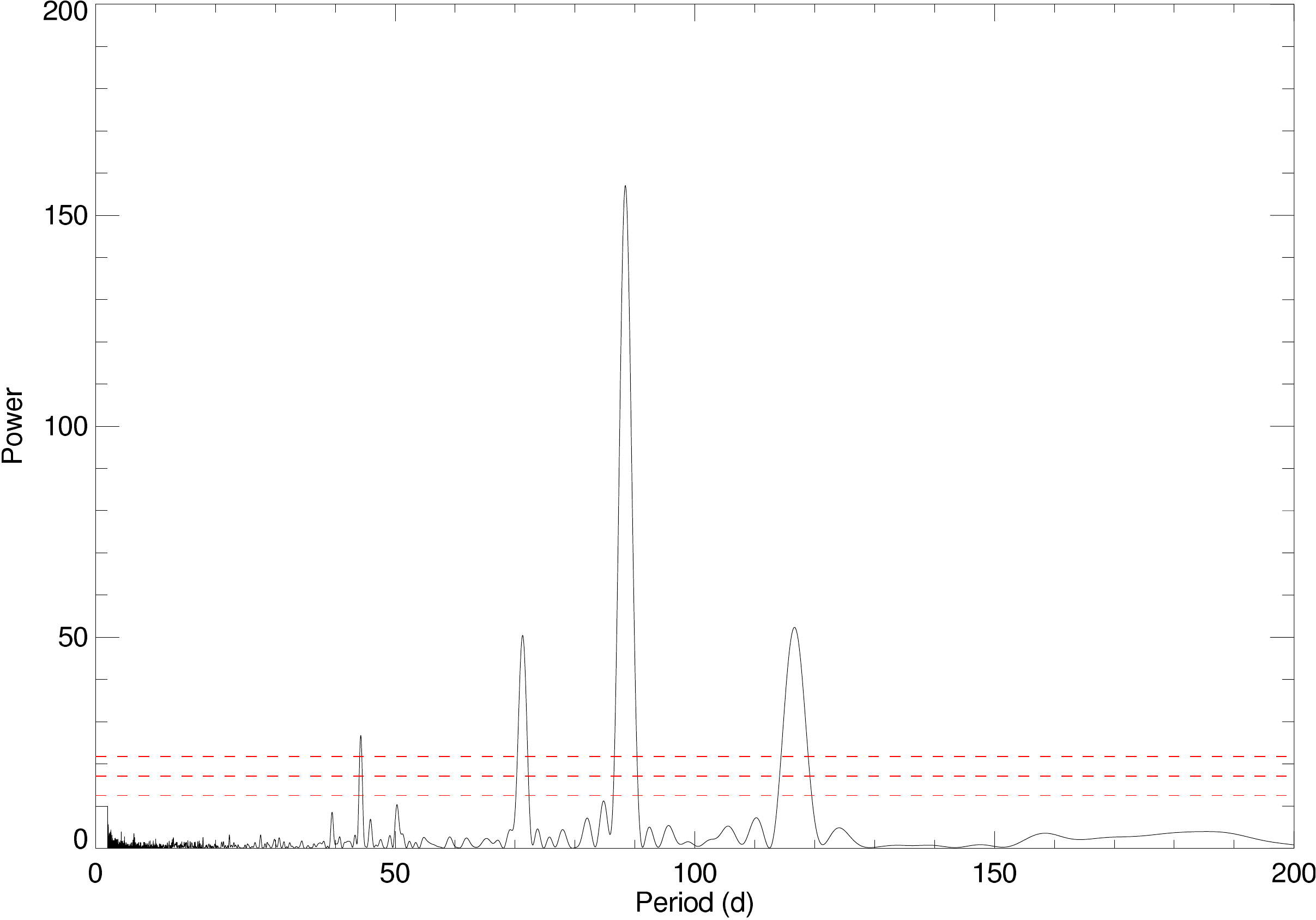}
 \caption[Power spectrum of the optical light curve of SXP91.1.]{Power spectrum
of the optical light curve of SXP91.1. Horizontal lines are 99\%, 99.99\% and
99.9999\% confidence levels. The peak period is $88.37\pm0.03$\,d. The second
harmonic is visible at 44.2\,d. The other peaks are likely to be aliasing caused
by the data gaps in the OGLE light curve.}
 \label{fig:91.1optls}
\end{figure}

\begin{figure*}
\hspace{-10pt}
 \includegraphics[scale=0.7]{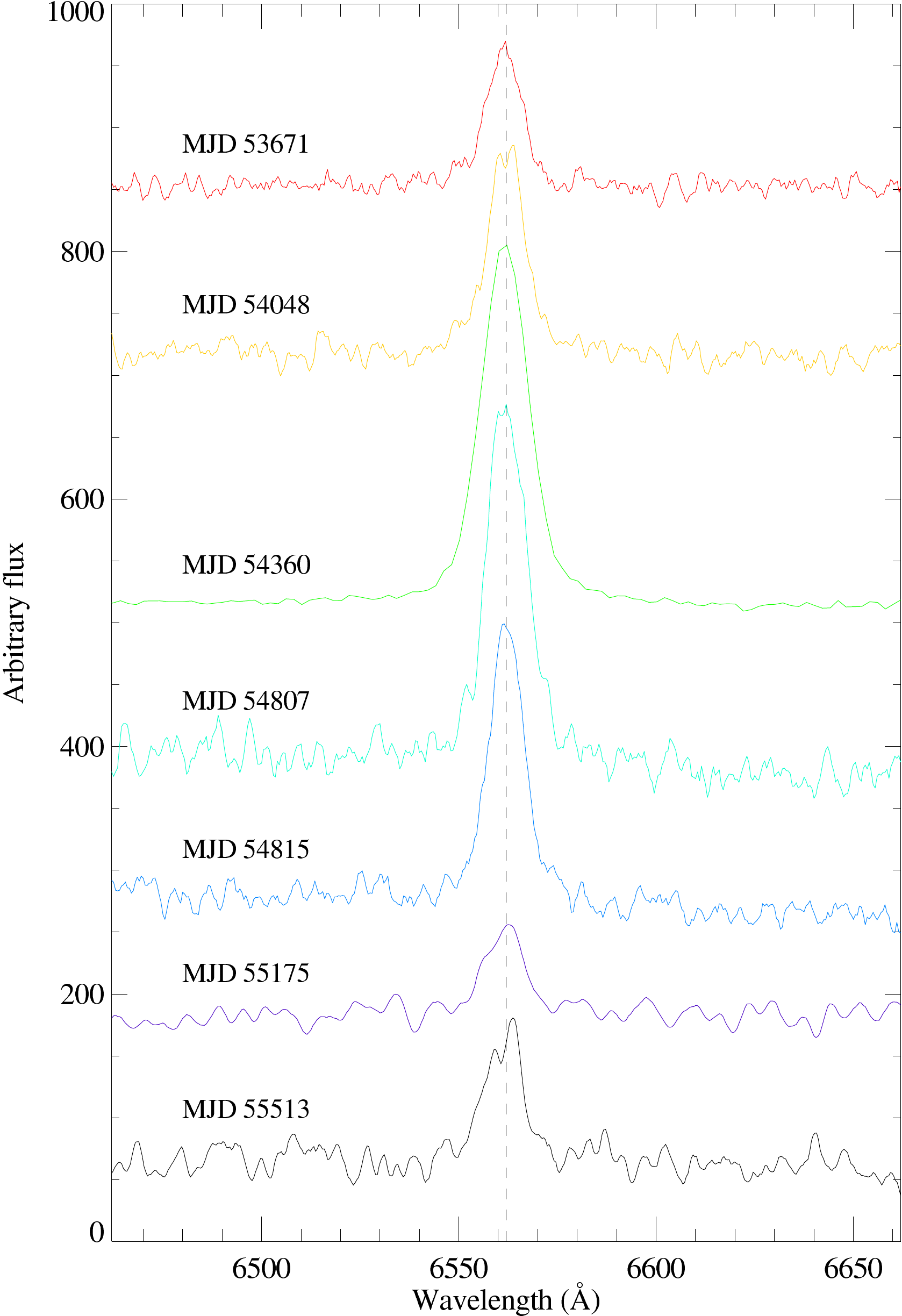}
 \caption[Optical spectra of SXP91.1 taken on the 1.9\,m telescope at SAAO and
3.6\,m telescope at La Silla.]{Optical spectra of SXP91.1 taken on the 1.9\,m
telescope at SAAO except the spectrum dated MJD 54360 which was taken on the
3.6\,m telescope at La Silla. The vertical line denotes the rest wavelength of
H$\alpha$. All spectra have been arbitrarily shifted in flux and smoothed with a
boxcar average of 5\,\AA{} width for viewing purposes.}
 \label{fig:91.1spec}
\end{figure*}

Figure \ref{fig:91.1optlc} shows the combined MACHO R-band and OGLE I-band light
curve of the optical counterpart to SXP91.1. The star is \#413 in the catalogue
of \citet{ma93} and has the OGLE III I.D. SMC 102.1.32. It shows clear periodic
variability, brightening by $\sim$0.06 magnitudes every 90 days, which is
undoubtedly due to the orbital motion of the neutron star. A Lomb-Scargle
analysis of the light curve reveals an extremely significant periodicity at
$88.37\pm0.03$\,d. The power spectrum is shown in Figure \ref{fig:91.1optls}.
This period is consistent with that found from analysis of the X-ray light curve
and the period originally reported in \citet{schmit04} and is thus confirmed as
the orbital period. There is very little long-term variability, besides a small
increase in brightness between MJD 52800 and 53800 by $\sim$\,0.05 magnitudes.
We note that the second of the 2 years of OGLE IV data is slightly fainter than
the first, suggesting a slight shrinking of the disc during the X-ray
outbursting phase. What is more interesting to note here, however, is that at
the time of maximum optical flux in the OGLE III part of the light curve (MJD
53000--53500) there is no X-ray emission detected from this source. However, in
the years preceding this small increase in optical flux, when the system appears
to be in a constant low state, there were three consecutive Type I X-ray
outbursts recorded in the \textit{RXTE} monitoring data (MJD 52300--52600). This
suggests X-ray emission only occurs, or at least only reaches the observer, at
times of optical minima. This could mean that when the disc is larger, it is
blocking most of the X-ray emission from the neutron star. This possibility is
discussed in detail later. Even in the low state, X-ray emission is not always
seen as is evident from the lack of X-ray activity between MJD 54000--55000.
Similar NIR photometric measurements to those described earlier were taken using
the IRSF telescope at SAAO. These J, H \& K images were taken on MJD 55179 and
allowed for a measurement of the magnitude of the source in these bands. Values
of J = $14.776\pm0.011$, H = $14.554\pm0.012$ \& K = $14.346\pm0.026$ suggest
the disc is in a similar state to the last 3 years of the OGLE III light curve
and consistent with the OGLE IV data. This implies that no dramatic change in
the circumstellar environment induced the current long series of X-ray
outbursts. Several H$\alpha$ spectra were also taken during this time period and
are shown in Figure \ref{fig:91.1spec}. The first five of the seven spectra
presented show almost constant equivalent widths (W$_{eq}$) of the H$\alpha$
line, with an average of -21.0\,\AA{} and range between -20.2 and -21.4\,\AA{}.
Errors on these measurements are between 0.7 and 1.1\,\AA{}. These spectra cover
the time period of the OGLE light curve, reinforcing the idea of a very stable
disc. The sixth and seventh spectrum, however, show equivalent widths of
-17.2$\pm$1.6 and -27.3$\pm$1.2\,\AA{} respectively. Both were taken after the
end of the OGLE light curve, suggesting there may have been some variability in
the disc emission after this time. The full \textit{RXTE} light curve for this
source is presented in the next section where the discussion of these
observations is continued.

\section{Discussion}

The utilisation of \textit{INTEGRAL} and \textit{XMM-Newton} to follow up on
unidentified \textit{RXTE} pulsars has proven to be very successful. In the
preceeding sections, we have discussed the discovery and localisation of a new
SMC pulsar (SXP175) to the HMXB candidate RX\,J0101.8--7223 and the association
of a suspected new pulsar with an already known one (SXP91.1). The cause of the
X-ray outburst from SXP175, which is the largest known from this HMXB, is not
clear as the state of the disc at the time of the outburst is not known. It is
likely that the disc increased in size, causing a small outburst. Previous
observations of the region by \textit{XMM-Newton} resulted in a positive
detection of the source on 5 occasions, each time being at a flux level
approximately 10\% of that detected in the ToO observation presented here
(Haberl, 2011, private communication). This not only explains why it has never
been seen with \textit{RXTE} at a significant level before, but reinforces the
idea that the disc may have grown and induced an outburst. The most recent of
the aforementioned \textit{XMM-Newton} detections was analysed by \citet{hep08}.
They present a power law fit to the spectrum that yields a photon index and
intrinsic SMC absorption that are almost identical to those values presented
here, despite the flux being $<$\,11\% of the flux measured in outburst. This
shows that there was no change in the accretion column or surrounding material
during the transition into outburst.

The analysis of data on SXP91.1 has shown that this particular BeXRB is quite
abnormal from the rest of the population. The following accounts a short summary
of the key observations made before potential explanations are given for the
peculiar behaviour of the pulsar:

\begin{itemize}
 \item it has a high intrinsic absorption column of
$(6.55\pm0.44)\times10^{22}$\,cm$^{-2}$
 \item it was shown to be emitting unpulsed X-ray emission at a luminosity
comparable to a Type I outburst ($8.2\times10^{35}$\,ergs\,s$^{-1}$) only one or
two orbits before the first detection of pulsations with \textit{RXTE}.
 \item it must have become much softer between the \textit{XMM-Newton} detection
and the \textit{INTEGRAL} detection to be seen in JEM-X and not IBIS. This could
mean it is softer in outburst (pulsed emission) but harder just before an
outburst or at a slightly lower flux level (unpulsed emission)
 \item the long-term light curve suggests a constant spin-up under phases of
outburst and quiescence. Quiescence is verified by the \textit{XMM-Newton}
non-detection with limiting luminosity of
$\sim$\,$4\times10^{33}$\,ergs\,s$^{-1}$
 \item its outbursting state seems not to be dependent on a change in the
circumstellar disc. X-ray emission is both detected and not detected, when the
optical flux (hence disc size) is unchanged. If anything, the X-ray outbursts
are seen at the lowest optical flux level. If the state of the disc does not
change and the orbital parameters do not change, then there is nothing to cause
accretion to start or stop. This may mean there is always X-ray emission but,
for some reason, this does not always reach the observer.
\end{itemize}

\begin{figure}
 \includegraphics[scale=0.35,angle=90]{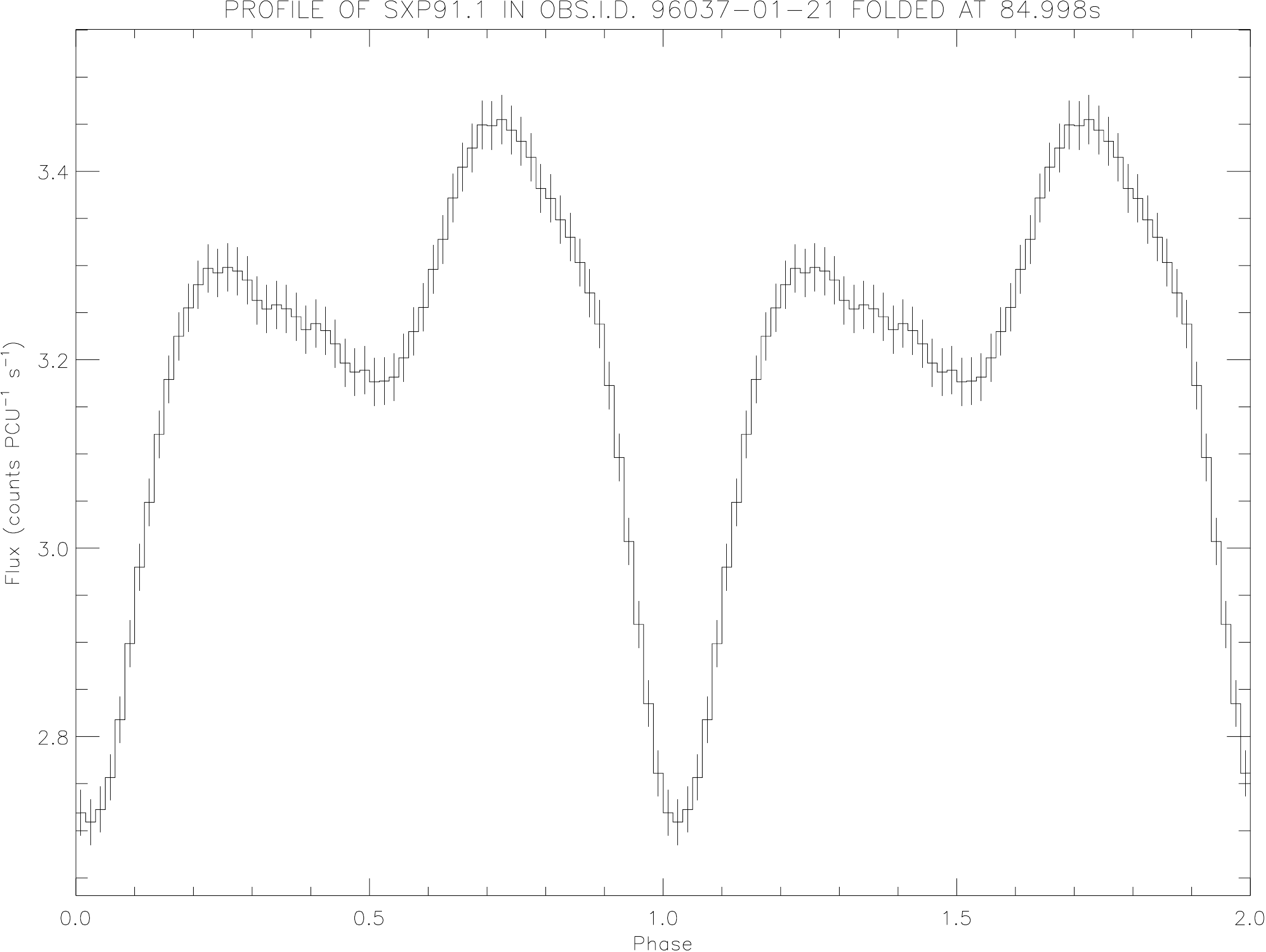}
  \caption[Example pulsed profile of SXP91.1.]{Example pulsed profile of SXP91.1
in the 3--10\,keV band, folded at the detected spin period. The profile has been
arbitrarily shifted in phase for viewing purposes.}
 \label{fig:sxp91.1pp}
\end{figure}

The \textit{XMM-Newton} observation on MJD 55101 falls 0.2 of an orbital phase
after periastron. Comparing this to the \textit{RXTE} folded light curve (see
Figure \ref{fig:sxp91.1pp}) suggests this lies very close to the point at which
pulsed emission ceases to be detected by \textit{RXTE}. This limit is known to
be around $10^{36}$\,ergs\,s$^{-1}$ in the SMC, which makes sense when the
\textit{XMM-Newton} derived luminosity is considered. The \textit{RXTE}
observation near MJD 55240 is the first confirmed detection of the source at a
luminosity of 6$\times$\,$10^{36}$\,ergs\,s$^{-1}$. Thus it seems that pulsed
X-ray emission may only begin between luminosities of
(0.8--6)\,$\times$\,$10^{36}$\,ergs\,s$^{-1}$. From the theory of magnetospheric
accretion (e.g. \citealt{corbet96}), it is highly unlikely that this lack of
pulsations at such a high luminosity is due to material not making its way to
the neutron star surface. More likely, it is due to changes in the absorbing
material around the system or changes in the hardness of the X-ray emission.
Spectral fits to the few \textit{RXTE} observations in which no other pulsar was
active seem to suggest a softer power law than the \textit{XMM-Newton}
observation. This makes the later \textit{INTEGRAL} detection with JEM-X more
understandable as the source may have softened beyond what is detectable with
IBIS, but it still does not explain why \textit{XMM-Newton} did not see
pulsations with sufficient counts and a normal value for the photon index.
Another explanation is an energy dependence of the pulsed profile. Since the
\textit{RXTE} and \textit{XMM-Newton} light curves were extracted in almost the
same energy ranges, the pulsed fraction must have increased at softer energies
between the \textit{XMM-Newton} and \textit{RXTE} detections. However, since
energies much above 10\,keV cannot be probed with XMM, this possibility is not
testable. There was marginal evidence for variability in the HI column density
between the \textit{XMM-Newton} and \textit{RXTE} observations, though this
parameter was much less well constrained in the fits than the photon index and
so drawing conclusions from this is not desirable.

The \textit{RXTE} pulsed profiles were studied to investigate variability in the
pulsed fraction and shape throughout the series of outbursts. The profile shown
in Figure \ref{fig:sxp91.1pp} is representative of most of the profiles
available. The pulsed shape is always double or triple peaked causing the
presence of multiple harmonics in the power spectrum, with substantial
variability in the relative strengths of the peaks. The pulsed fraction is
approximately constant between outbursts, between 0.1--0.2, but shows a strong
correlation with pulsed flux (and hence time) during individual outbursts. This
may suggest an almost constant unpulsed component and that the pulsed component
is much more variable under changing accretion rates. Again, it is hard to
quantify the level of background emission in \textit{RXTE} data, so these pulsed
fractions are lower limits. To assess the non-pulsed detection of
SXP91.1 during XMM observation 0601210701 we constructed a number of simulated
light curves based upon the actual XMM light curve.  A sinusoidal modulation
with period 85.4~s was constructed with same sampling of the XMM light curve
this was then combined with the actual X-ray flux measurement such that the
average flux remained the same and that a specific pulse fraction (PF) was
achieved: PF$ = \frac{F_{max}-F{min}}{F_{max}+F{min}}$.  Each flux value is then
shifted based upon the original error bar and assuming Gaussian errors.  Two
thousand light curves were simulated for each PF  over a range of 0.00--0.25 in
steps of 0.01.  The Lomb-Scargle periodogram was calculated for each light curve
and all those corresponding to a given PF were averaged. To detect a signal in
the data with a significance of $>5\sigma$ would require a pulse fraction of
$>14.8\%$ and at $>3\sigma$ would require a pulse fraction of $>11.1\%$. Our
simulation ties in nicely with the \textit{RXTE} observations, in that it shows
the pulsed fraction of the source must have increased from below around 10\%
prior to the \textit{RXTE} observations, to above this value during them,
allowing for the spin period detections.

\begin{figure}
 \includegraphics[scale=0.35,angle=90]{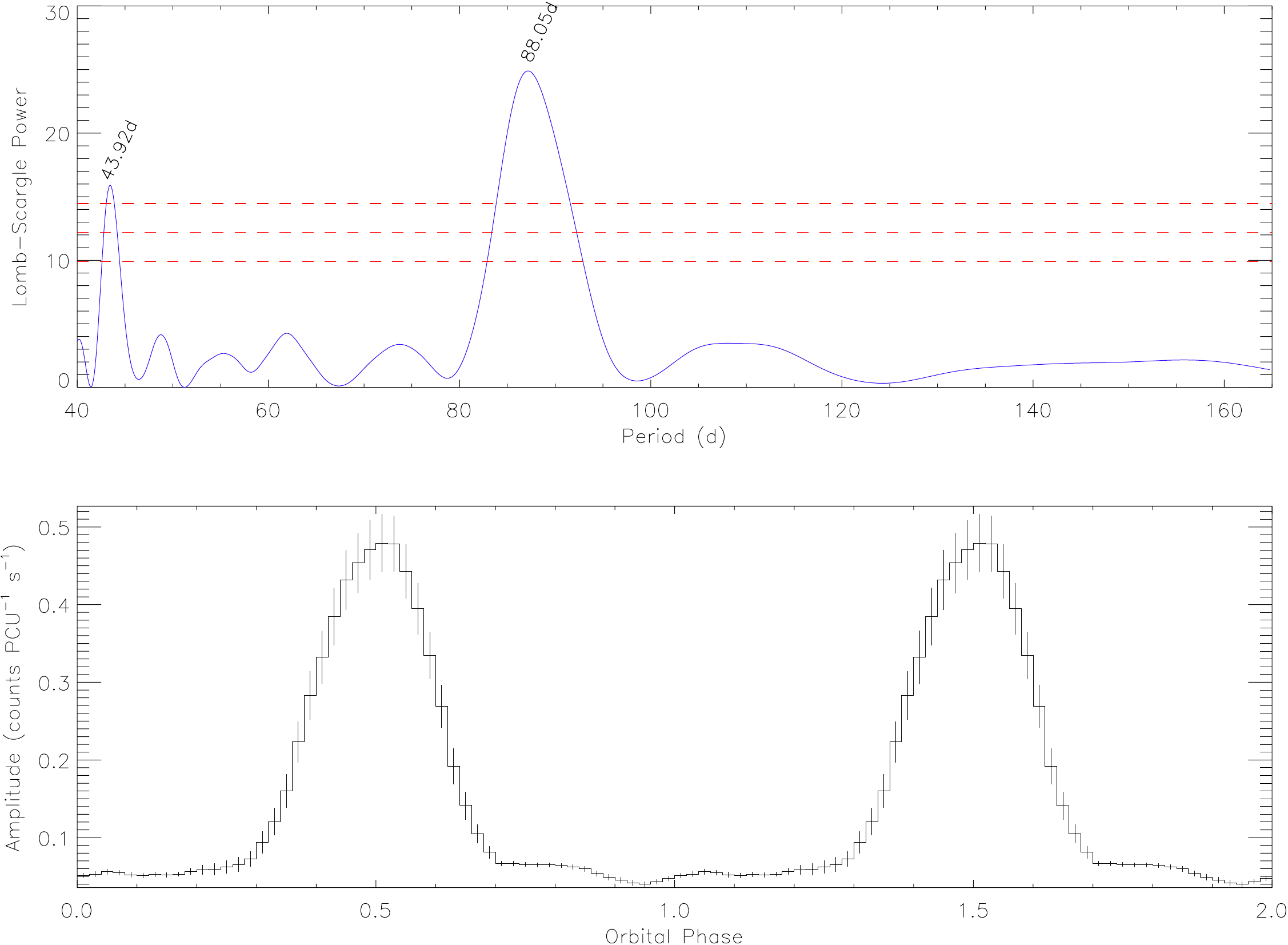}
  \caption[Power spectrum and folded X-ray light curve of SXP91.1.]{Top: orbital
power spectrum of the long-term X-ray light curve of SXP91.1. The peak at
88.1\,d signifies the orbital period of the system and agrees well with the
optically derived period. The horizontal lines are 99\%, 99.99\% and 99.9999\%
significances. Bottom: the X-ray light curve folded on the orbital period. The
ephemeris of maximum amplitude is MJD 55516.43$\pm$2.64. Increased flux is
apparent over a phase space of 0.4, higher than most other BeXRBs.}
 \label{fig:orb1}
\end{figure}

\begin{figure}
\hspace{10pt}
 \includegraphics[scale=0.35,angle=90]{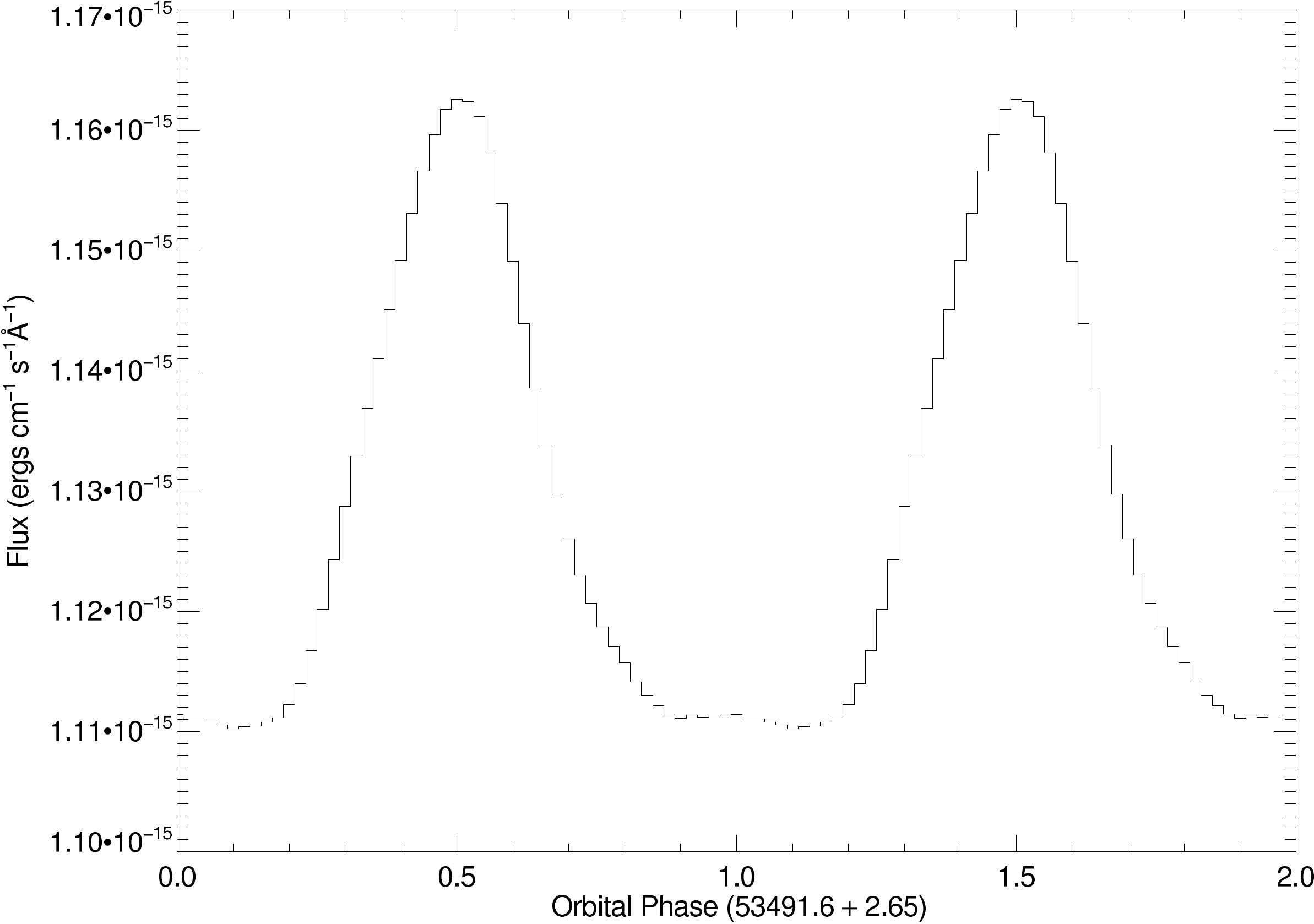}
  \caption[Folded optical light curve of SXP91.1.]{Folded optical light curve of
SXP91.1. As with the X-ray light curve, a broad peak is observed suggesting the
disturbance in the disc lasts for almost half an orbit.}
 \label{fig:orb2}
\end{figure}

Figures \ref{fig:orb1} and \ref{fig:orb2} show the folded orbital X-ray and
optical light curves respectively. Both show a broad profile not in-keeping with
the fast-rise, exponential decay shape that is usually present in eccentric
accreting binaries. This broad emission region, spanning around half an orbital
period, is indirect evidence of a low eccentricity in this system as the neutron
star is accreting far beyond periastron. If true, this observation places
SXP91.1 into a minority group amongst BeXRB pulsars. This may also help to
explain the unusual spin-up that is observed through phases of quiescence in
this system. If the orbit is near-circular, it could be accreting at a very low
level all the time, spinning the neutron star up, whilst only becoming an X-ray
pulsar when the accretion rate increases. However, as mentioned earlier, the
cause of such an increase is hard to understand given the stability of the
circumstellar disc. The optically and X-ray derived ephemerides agree well, with
a suggestion that the peak in optical flux may lag the X-ray peak by no more
than 6 days. The errors are large, however, so this value could be much less.
Should the lag be real (5--10\% of an orbital period), it reinforces what is
currently thought to occur during periastron passage; that the disc only begins
to be disturbed after the neutron star passes periastron \citet{jones12}.

The constant state of the circumstellar disc is unusual during such a dramatic
change in X-ray activity. NIR fluxes suggest that the constant nature of the
disc seen in the OGLE light curve has continued well into the current X-ray
active phase as discussed above. H$\alpha$ spectra taken before and during the
X-ray active phase (see Figure \ref{fig:91.1spec}) shed some light on the
problem, but also raise some further issues. The H$\alpha$ equivalent width
measurement on MJD 55175 suggests a small decrease in the disc emission at the
onset of the X-ray outburst, whilst the following measurement is evidence of a
significant increase in the disc emission. At first, this may appear as evidence
for the growth of the disc many hundreds of days into the series of X-ray
outbursts. However, on closer inspection this observation falls nearly 2 weeks
before periastron and so could just have been induced by the interacting neutron
star rather than being an intrinsic change in the disc. A counter argument to
this though, is that four of the other spectra were taken as close or closer to
periastron than this one and show no sign of an increase in flux. If the flux
increase is intrinsic to the disc, it could indicate a significant lag between
the start of the X-ray outburst and increased optical activity. Beyond this
variability, these H$\alpha$ spectra show a further intriguing feature: The flux
and shape of the profiles may correlate with orbital phase. The 3 profiles with
the smallest W$_{eq}$ were taken within a few days of periastron, whilst the 3
with the largest W$_{eq}$ were taken at least 2 weeks before or after
periastron. This could be evidence of the neutron star truncating the disc
slightly \citep{neg01}, though the reason for the changing profile shapes and
their orbital phase distribution is very hard to understand with the data
available.

\subsection{Investigating the spin-up of SXP91.1}

To explore the apparent spin-up of SXP91.1 we assessed the characteristics and
data quality of each of the \textit{RXTE} observations. The dataset was filtered
to include only observations where a periodic signal between 85--95\,s, was
detected at a significance of $>$\,99.99\% and which had a collimator response
of $>$\,0.4. One additional data-point was excluded as clearly being an outlier
with a $\sim$\,84s period at early epochs. This left 37 \textit{RXTE}
observations of SXP91.1 where a periodic signal was detected. Errors on the
periods were initially calulated using the analytical equations of
\citet{horne86}. However, this method relies on well behaved, high quality data
(e.g. the data shows simple white noise). As the strength of the the source
signal varies significantly between observations we independently assessed the
period errors for the Lomb-Scargle technique using a bootstrap method (see e.g.,
\citealt{1995GeoRL..22..307K}).

For each of the filtered observations a simulated light curve is generated by a
{\it bootstrap with replacement} technique which randomly selects a subset of
$\sim$68\% of the original light curve points.  The corresponding Lomb-Scargle
periodogram is calculated and the period of the peak of highest power is
recorded; this is repeated 200,000 times.  To protect against extreme outliers
in the simulations periods more than 3$\sigma$ away from the median of the
simulated periods are excluded, the standard deviation of the filtered sample of
simulations is then taken as an estimate of the error.

\begin{figure}
\hspace{-10pt}
\includegraphics[scale=0.55]{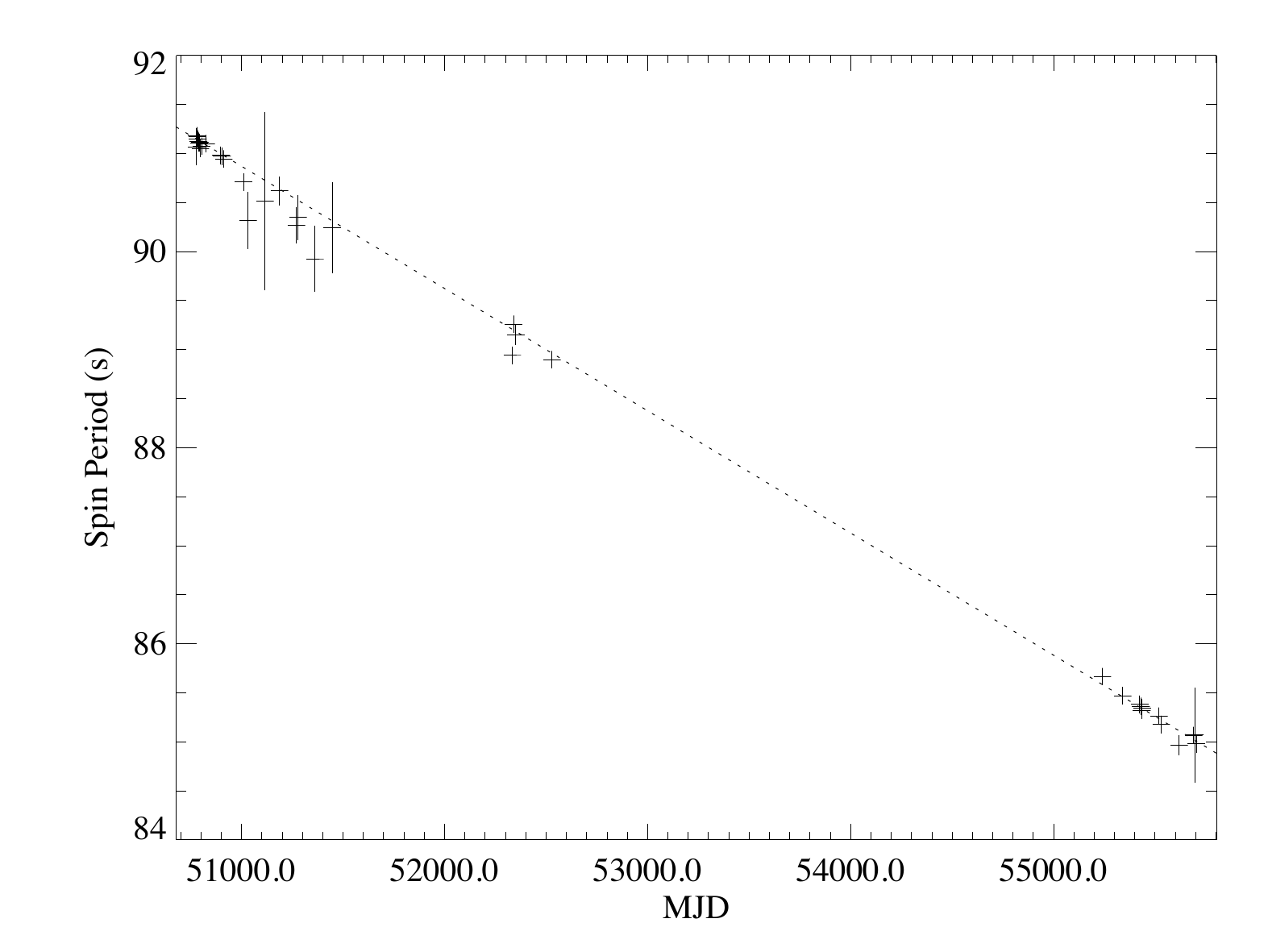}
\caption{The fitted linear spin-up model of SXP91.1 is plotted as a dashed line
over the RXTE spin-period detections.  The individual RXTE period measurements
are plotted with 1$\sigma$ errors derived from bootstrapping the original
dataset, see text for details.}
\label{fig:spin-up}
\end{figure}

In all except two of the observations the bootstrap error was comparable to or
greater than the errors from \citet{horne86} and so to be conservative were
adopted for the spin-up analysis.  The remaining two observations were during a
bright outburst and the boostrap re-sampling did not yield distribution of
periods with a finite width and hence for these two observations we adopt the
analytical errors.  The observations span $\sim$13.5 years, applying a simple
linear spin-up model over all observations of:

\begin{equation}
 P(t) = P(t_{0}) + \dot{P}(t - t_{0})
  \label{equ:spin}
\end{equation}

\noindent yields a good quality fit to the data with $\chi ^{2}/\mathrm{dof} =
41.2/35 = 1.18$ and giving a spin-up, $\dot{P} = 1.442 \pm 0.005 \times 10^{-8}
ss^{-1}$. We attempted to incorporate the orbital modulation of the pulsar into
the spin-up model however the sampling and quality of the data were not
sufficient to constrain fitting a model of this type and consequently the simple
linear spin-up model was found to be the best representation of the data.  The
data points with the bootstrap errors and best fit linear model are shown in
Fig.~\ref{fig:spin-up}.

Using this measurement, the average X-ray luminosity can be estimated assuming a
given neutron star mass and magnetic moment, and assuming a constant accretion
rate throughout our 13 years of observations. This final assumption is not
unreasonable given the good fit to the spin periods presented above. Using the
accretion models of \citep{gl79} whereby the compact object accretes
from a disc, the spin-up and spin period measurements give an average
luminosity of ($1.9\pm0.2$)$\times10^{37}$\,ergs\,s$^{-1}$. This is markedly
different to the $8.2\times10^{35}$\,ergs\,s$^{-1}$ calculated from the
\textit{XMM-Newton} spectrum shown earlier and hints at two or three possible
explanations: Firstly, the X-rays produced by the neutron star are in
the region of 
$10^{37}$\,ergs\,s$^{-1}$ predicted by the model, but this light has been
attenuated 
by an absorbing material around the neutron star or the binary system.
Secondly, 
our interpretation of a constant accretion rate and spin
up are correct, and the over-estimated average luminosity is due to our
assumption of the mass or magnetic moment of the neutron star. In this case the
mass of the neutron star would be less than the 1.4$M_{\odot}$ assumed or the
magnetic moment would have to be higher than the 4.8$\times10^{29}$gauss
cm$^{3}$ assumed. For completeness, we have calculated the possible
extreme values of B and M
assuming the observed average luminosity as an input into the accretion model. 
When calculating B, we used the observed L, P and $\dot{P}$ and assumed a radius
and mass of 10$^{6}$cm and 1.4$M_\odot$ respectively. When calculating M, we
assumed the above 
values and a magnetic moment of 4.8$\times10^{29}$gauss cm$^{3}$. The derived
value of the
magnetic field is $(3.1\pm0.1)\times10^{13}$G, consistent with a higher field
than assumed in our luminosity calculation. 
The value of M was not constrained by this calculation, suggesting that the low
assumed magnetic field is indeed impossible for this observed luminosity. 
A third and final possibility is that our constant spin-up assumption could be
incorrect and the
neutron star has undergone varying levels of interaction with the Be star disc
resulting in a variable luminosity. The latter scenario seems less likely given
the evidence we have seen from the spin-up fit and the optical light curves.
Thus, it is more likely that either the neutron star in this system 
has a larger magnetic field than average, or there is a
variable absorbing material 
around the neutron star. There is strong evidence that such a variable absorber
exists as 
it seems the only way to produce the variability seen in the X-rays without
changing the 
accretion rate and hence, the spin up of the neutron star. A potential such
absorber may be
a disc that absorbs the X-ray emission at certain times, possibly when it is
enlarged.

\section{Conclusions}

This paper has presented the discovery of a new 175\,s pulsar in the SMC and its
association with the HMXB candidate RX\,J0101.8--7223. This system seems to be a
very standard BeXRB system that has shown low levels of X-ray activity fairly
regularly and a rare larger outburst that allowed the pulse period to be found.
The large outburst is likely caused by an increase in the size of the
circumstellar disc around the counterpart that, until now, has been very stable.
Analysis of the optical light curve allowed an estimate of the orbital period of
the system of $87.2\pm0.2$\,d. An archival spectrum has allowed the counterpart
to be classified as a B\,0--B\,0.5 IIIe star, similar to most other known
BeXRBs.

SXP91.1 has proven to be very different to SXP175, with very regular outbursts,
large modulation in its optical light curve and peculiar spin-up and X-ray
emission characteristics. We have endeavoured to explain these aspects by
combining all available X-ray, optical and IR data. It seems clear that the
neutron star has always, and is continuing to, interact with the circumstellar
disc at every periastron passage from the periodic variations in the OGLE light
curve, though this interaction does not always lead to observable X-ray
emission, pulsed or otherwise, reaching the observer. The idea of constant
accretion in this system, despite the lack of X-ray emission is backed up by the
relatively constant spin-up of the neutron star throughout the entire
\textit{RXTE} monitoring campaign. Since the lack of pulsations at such a high
flux is improbable for this spin period, the most likely explanation from the
data presented here is that either a nearby absorbing material sometimes acts to
prevent the X-ray emission reaching the observer, or a change in the geometry of
the emission column or magnetic field of the neutron star directs the emission
away from us at certain times. The other possible scenario is that a
near-circular orbit barely brings the neutron star into contact with the disc,
causing a weak interaction every periastron passage. This could spin the neutron
star up and cause a spike in the optical light curve without resulting in a
detectable X-ray outburst. At the times when an outburst is seen, the accretion
rate is likely to have increased, though with a very stable disc it is unclear
what could cause the system to outburst. Determining the orbit of this system
would provide a huge step to understanding its strange behaviour, though this
would have to be done through optical radial velocity work, as the X-ray
outbursts are not suitable for such an analysis.

\section*{Acknowledgements}

LJT is supported by a Mayflower scholarship from the University of Southampton.
We are grateful to the staff at SAAO, and Tetsuya Nagata, for support during the
1.9m and IRSF telescope runs. Based on observations made with ESO Telescopes at
the La Silla Observatory under programme ID 079.D-0371(A). ABH acknowledges that
this research was supported by a Marie Curie International Outgoing Fellowship
within the 7th European Community Framework Programme (FP7/2007--2013) under
grant agreement no. 275861. We thank Helen Klus for her help with some of the
data analysis and the referee for their comments on which this paper has
benefitted.

\bsp

\label{lastpage}


\begin{thebibliography}{99}

\bibitem[\protect\citeauthoryear{Antoniou et al.}{2009}]{ant09}
Antoniou V., Hatzidimitriou D., Zezas A., Reig P., 2009, ApJ, 707, 1080
\bibitem[\protect\citeauthoryear{Coe et al.}{2005}]{coe05}
Coe M.J., Edge W.R.T., Galache J.L., McBride V.A., 2005, MNRAS, 356, 502
\bibitem[\protect\citeauthoryear{Corbet}{1996}]{corbet96}
Corbet R.H.D., 1996, ApJ, 457, L31
\bibitem[\protect\citeauthoryear{Corbet et al.}{1998}]{cor98}
Corbet R., Marshall F.E., Lochner J.C., Ozaki M., Ueda Y., 1998, IAU Circ., 6803
\bibitem[\protect\citeauthoryear{Corbet et al.}{2010}]{cor10}
Corbet R.H.D., Bartlett E.S., Coe M.J., McBride V.A., Townsend L.J., Schurch
M.P.E., Marshall F.E., 2010, Astron. Telegram, 2813
\bibitem[\protect\citeauthoryear{Diaz \& Bekki}{2011}]{diaz11}
Diaz J., Bekki K., 2011, MNRAS, 413, 2015
\bibitem[\protect\citeauthoryear{Dickey \& Lockman,}{1990}]{dic90}
Dickey J. M. \& Lockman F. J., 1990, ARA\&A, 28, 215
\bibitem[\protect\citeauthoryear{Dray}{2006}]{dray06}
Dray L.M., 2006, MNRAS, 370, 2079
\bibitem[\protect\citeauthoryear{Evans et al.}{2004}]{evans04}
Evans C.J., Howarth I.D., Irwin M.J., Burnley A.W., Harries T.J., 2004, MNRAS,
353, 601
\bibitem[\protect\citeauthoryear{Galache et al.}{2008}]{gal08}
Galache J.L., Corbet R.H.D., Coe M.J., Laycock, S., Schurch M.P.E., Markwardt
C., Marshall F.E., Lochner, J., 2008, ApJS, 177, 189
\bibitem[\protect\citeauthoryear{Gardiner \& Noguchi}{1996}]{gard96}
Gardiner L.T., Noguchi M., 1996, MNRAS, 278, 191
\bibitem[\protect\citeauthoryear{Ghosh \& Lamb}{1979}]{gl79}
Ghosh P. \& Lamb F.K., 1979, ApJ, 234, 296
\bibitem[\protect\citeauthoryear{Goldwurm et al.}{2003}]{gold03}
Goldwurm A., et al., 2003, A\&A, 411, 223
\bibitem[\protect\citeauthoryear{Haberl \& Sasaki}{2000}]{hab00}
Haberl F., Sasaki M., 2000, A\&A, 359, 573
\bibitem[\protect\citeauthoryear{Haberl, Eger \& Pietsch}{2008}]{hep08}
Haberl F., Eger P., Pietsch W., 2008, A\&A, 489, 327
\bibitem[\protect\citeauthoryear{Hilditch, Howarth \&
Harries}{2005}]{hilditch05}
Hilditch R.W., Howarth I.D., Harries T.J., 2005, MNRAS, 357, 304
\bibitem[\protect\citeauthoryear{Hill et al.}{2010}]{hill10}
Hill A.B., Dubois R., Torres D.F., on behalf of the Fermi-LAT collaboration,
2010, To be published in the book of proceedings of the 1st Sant Cugat Forum on
Astrophysics, "ICREA Workshop on the high-energy emission from pulsars and their
systems", held in April, 2010, arXiv:1008.4762
\bibitem[\protect\citeauthoryear{Horne \& Baliunas}{1986}]{horne86}
Horne J.H., Baliunas S.L., 1986, ApJ, 302, 757
\bibitem[\protect\citeauthoryear{Harries, Hilditch \& Howarth}{2003}]{hhh03}
Harries T.J., Hilditch R.W., Howarth I.D., 2003, MNRAS, 339, 157
\bibitem[\protect\citeauthoryear{Jahoda et al.}{2006}]{jahoda06}
Jahoda K., Markwardt C. B., Radeva Y., Rots A. H., Stark M. J.,
Swank J. H., Strohmayer T. E., Zhang W., 2006, ApJS, 163, 401.
\bibitem[\protect\citeauthoryear{Jones et al. in prep.}{2012}]{jones12}
Jones J., Coe M.J., et al., 2012, in prep.
\bibitem[\protect\citeauthoryear{Kato et al.}{2007}]{kato07}
Kato D. et al., 2007 PASJ 59, 615.
\bibitem[\protect\citeauthoryear{Kawano \& Higuchi}{1995}]{1995GeoRL..22..307K}
Kawano H., Higuchi T., 1995, GeoRL, 22, 307
\bibitem[\protect\citeauthoryear{Lebrun et al.}{2003}]{lebrun03}
Lebrun F., et al., 2003, A\&A, 411, 141
\bibitem[\protect\citeauthoryear{Lennon}{1997}]{len97}
Lennon D.J., 1997, A\&A, 317, 871
\bibitem[\protect\citeauthoryear{Lund et al.}{2003}]{lund03}
Lund N., et al., 2003, A\&A, 411, 231
\bibitem[\protect\citeauthoryear{Marshall et al.}{1997}]{marshall97}
Marshall F.E., Lochner J.C., Takeshima T., 1997, IAU Circ., 6777, 2
\bibitem[\protect\citeauthoryear{McBride et al.}{2008}]{mcbride08}
McBride V.A., Coe M.J., Negueruela I., Schurch M.P.E., McGowan K.E., 2008,
MNRAS, 388, 1198
\bibitem[\protect\citeauthoryear{Meyssonnier \& Azzopardi}{1993}]{ma93}
Meyssonnier N., Azzopardi M., 1993, A\&AS, 102, 451
\bibitem[\protect\citeauthoryear{Morgan, Keenan \& Kellman}{1943}]{mkk43}
Morgan W.W., Keenan P.C., Kellman E., 1943, An Atlas of Stellar Spectra, with an
Outline of Spectral classification. Univ. Chicago Press, Chicago
\bibitem[\protect\citeauthoryear{Nagashima et al.}{1999}]{nag99}
Nagashima C., et al. 1999, in Star formation 1999, ed. T. Nakamoto, (Nobeyama
Radio Observatory), 397
\bibitem[\protect\citeauthoryear{Negueruela \& Okazaki}{2000}]{neg00}
Negueruela I., Okazaki A.T., 2000, ASPC, 214, 713
\bibitem[\protect\citeauthoryear{Negueruela \& Okazaki}{2001}]{neg01}
Negueruela I., Okazaki A.T., 2001, A\&A, 369, 108
\bibitem[\protect\citeauthoryear{Putman et al.}{2003}]{putman2003}
Putman M.E., Staveley-Smith L., Freeman K.C., Gibson B.K., Barnes D.G., 2003,
ApJ, 586, 170
\bibitem[\protect\citeauthoryear{Reig, Fabregat \& Coe}{1997}]{reig97}
Reig P., Fabregat J., Coe M.J., 1997, A\&A, 322, 193
\bibitem[\protect\citeauthoryear{Sasaki, Haberl \& Pietsch}{2000}]{sasaki00}
Sasaki M., Haberl F., Pietsch W., 2000, A\&AS, 147, 75
\bibitem[\protect\citeauthoryear{Scaringi et al.}{2010}]{scaringi2010}
Scaringi S., et al., 2010, A\&A, 516, 75
\bibitem[\protect\citeauthoryear{Schmidtke et al.}{2004}]{schmit04}
Schmidtke P.C., Cowley A.P., Levenson L., Sweet K., 2004, AJ, 127, 3388
\bibitem[\protect\citeauthoryear{Schmidtke \& Cowley}{2006a}]{schmit06}
Schmidtke P.C., Cowley A.P., 2006, AJ, 132, 919
\bibitem[\protect\citeauthoryear{Schurch et al.}{2011}]{schurch10}
Schurch M.P.E., Coe M.J., McBride V.A., Townsend L.J., Udalski A., Haberl F.,
Corbet R.H.D., 2011, MNRAS, 412, 391
\bibitem[\protect\citeauthoryear{Sguera et al.}{2005}]{sgu05}
Sguera V., et al., 2005, A\&A, 444, 221
\bibitem[\protect\citeauthoryear{Sidoli}{2011}]{sid11}
Sidoli L., 2011, Invited talk at the 25th Texas Symposium on Relativistic
Astrophysics, held in Heidelberg, Germany, on December 6-10, 2010,
arXiv1103.6174
\bibitem[\protect\citeauthoryear{Struder et al.}{2001}]{struder01}
Str{\"u}der L., et~al. 2001, A\&A, 365, L18.
\bibitem[\protect\citeauthoryear{Townsend et al.}{2011b}]{town11b}
Townsend L.J., Drave S.P., Corbet R.H.D., Coe M.J., Bird A.J., 2011, Astron.
Telegram, 3311.
\bibitem[\protect\citeauthoryear{Turner et al}{2001}]{turner01}
Turner M.~J.~L., et~al. 2001, A\&A, 365, L27.
\bibitem[\protect\citeauthoryear{Ubertini et al.}{2003}]{ubertini03}
Ubertini P., et al., 2003, A\&A, 411, 131
\bibitem[\protect\citeauthoryear{Udalski}{2008}]{uda08}
Udalski A., 2008, Acta Astron., 58, 187
\bibitem[\protect\citeauthoryear{Walborn \& Fitzpatrick}{1990}]{wal90}
Walborn N.R., Fitzpatrick E.L., 1990, PASP, 102, 379
\bibitem[\protect\citeauthoryear{Wegner}{2006}]{weg06}
Wegner W., 2006, MNRAS, 371, 185
\bibitem[\protect\citeauthoryear{Winkler et al.}{2003}]{winkler03}
Winkler C., et al., 2003, A\&A, 411, 1
\bibitem[\protect\citeauthoryear{Yokogawa et al.}{2003}]{yoko03}
Yokogawa J., Imanishi K., Tsujimoto M., Koyama K., Nishiuchi M., 2003, PASJ, 55,
161


\end{thebibliography}
\end{document}